\documentclass{aa}
\usepackage{natbib}
\bibpunct{(}{)}{;}{a}{}{,} 
\defcitealias{ZanFer09}{Paper I}
\usepackage[varg]{txfonts}
\usepackage{graphicx}

\begin{document}
   \title{MHD simulations of accretion onto a dipolar magnetosphere}

   \subtitle{II. Magnetospheric ejections and stellar spin-down}

   \author{C. Zanni
          \inst{1}
          \and
          J. Ferreira
          \inst{2}          
          }

   \offprints{C. Zanni, \email{zanni@oato.inaf.it}}

   \institute{INAF - Osservatorio Astrofisico di Torino, Strada Osservatorio 20, 10025, Pino Torinese, Italy \and
                    UJF-Grenoble 1/CNRS-INSU,
                    Institut de Plan\'etologie et d'Astrophysique de Grenoble (IPAG) UMR 5274,
                   Grenoble, F-38041, France
             }

   \date{Received ..................... ; accepted ................... }

 
  \abstract
   {} 
   {This paper examines the outflows associated with the interaction of a stellar magnetosphere with an accretion disk. 
     In particular, we investigate the \emph{magnetospheric ejections} (MEs) due to the expansion and reconnection of the
     field lines connecting the star with the disk. Our aim is to study the dynamical properties of the outflows and evaluate their 
     impact on the angular momentum evolution of young protostars.}
   {Our models are based on axisymmetric time-dependent magneto-hydrodynamic simulations of the interaction of the dipolar magnetosphere of a
     rotating protostar with a viscous and resistive disk, using alpha prescriptions for the transport coefficients. Our simulations are 
     designed in order to model: the accretion process and the formation of accretion funnels; the periodic inflation/reconnection of the 
     magnetosphere and the associated MEs; the stellar wind.}
   {Similarly to a magnetic slingshot, MEs can be powered by the rotation of both the disk and the star so that they can efficiently remove angular momentum
    from both. Depending on the accretion rate, MEs can extract a relevant fraction of the accretion torque and, together with a weak but
    non-negligible stellar wind torque, can balance the spin-up due to accretion. When the disk truncation approaches the corotation radius,
    the system enters a ``propeller'' regime, where the torques exerted by the disk and the MEs can even balance the spin-up due to the stellar
    contraction.}
   {Magnetospheric ejections can play an important role in the stellar spin evolution. Their spin-down efficiency can be compared to other 
     scenarios, such as the \citeauthor{Ghosh79}, X-wind or stellar wind models. Nevertheless, for all scenarios, an efficient spin-down torque requires
     a rather strong dipolar component, which has been seldom observed in classical T Tauri stars. A better analysis of the torques acting on
     the protostar must take into account non-axisymmetric and multipolar magnetic components consistent with observations.}
     

   \maketitle
%
%

\section{Introduction}
\label{sec:intro}

Classical T Tauri stars (CTTS) are pre-main sequence stars showing clear signatures of accretion from a surrounding
accretion disk \citep{Edwards94,Hartmann98} and ejection in the form of collimated jets \citep{Cabrit90,Burrows96}.
The evolution of their rotation period represents an interesting puzzle. As soon as they become visible after the Class 0-I
embedded phases, a relevant fraction of CTTS appears to rotate well below their break-up limit, 
with rotation periods around 1-10 days \citep{Bouvier93}. Besides, their rotation rate appears to be rather constant 
during the accreting evolutionary phases lasting a few million years \citep{Irwin09}. On the other hand, these protostars 
are still actively accreting and contracting so that they would be expected to spin-up at break-up in $\sim 10^6$ years. 
Clearly, CTTS require an efficient spin-down mechanism to explain their rotational evolution.

Since CTTS are known to be magnetically active \citep[see e.g.][]{JK07, Yang11}, different magneto-hydrodynamic (MHD) 
mechanisms of angular momentum removal have been proposed.
In the \citet{Ghosh79} model, originally developed for pulsars,  the disk itself extracts angular momentum from the star
along the field lines connecting the star with the disk in the region beyond the corotation radius, where the disk
rotates slower than the star. On the other hand, it has been shown that the efficiency of the \citeauthor{Ghosh79} 
mechanism is drastically reduced because of the limited size of the connected magnetosphere \citep{Matt05a}
and the dilution of the poloidal field beyond the corotation radius \citep[][hereafter Paper I]{AgaPap00,ZanFer09}.

Other solutions are based on the presence of outflows, drawing out angular momentum from the star-disk system, instead 
of transfering it back to the disk.
\citet{Shu94} proposed that an ``X-Wind'' launched along the open stellar magnetic surfaces threading the disk around
corotation can extract a substantial amount of angular momentum from the disk before it is transferred to the star, 
so as to cancel at least the spin-up torque due to accretion. Even if models of wide-angle X-winds are feasible 
\citep{Anderson05, Cai08}, a fully self-consistent calculation of the disk-outflow dynamical connection is currently missing.

\citet{Ferreira00} investigated a different magnetic configuration, where a magnetic neutral line is formed at the star-disk 
interface due to the cancellation of the stellar dipolar field by the disk field. Such a reconnection site has been envisioned 
to drive massive unsteady ejection events, mainly powered by the stellar rotation. These "Reconnection X-winds" provide a 
very efficient spinning down mechanism for early low-mass protostars (Class 0 and I objects), able to brake a maximally 
rotating initial core down to observed values. However this model has not been designed for CTTS and requires a specific 
magnetic topology, not addressed here.

Stellar winds provide a spin-down torque extracting angular momentum along the open magnetospheric field lines 
anchored onto the stellar surface \citep{Matt05b, Sauty11}. \citet{Matt08b} estimated 
that the wind mass flux is likely to be of the order of $\sim 10\%$ of the accretion rate in order to balance at least the torque
due to accretion. These stellar winds would carry the entire mass flux typically observed in T Tauri jets, 
which seems unlikely \citep{Cabrit09}. Indeed, such a high ejection efficiency presents a 
serious energetic problem \citep{Ferreira06}: since CTTS are slow rotators and their centrifugal push is not 
strong enough to drive these outflows, an extra energy input is required. \citet{Matt05b} proposed that it 
could come from the accretion power carried in onto the star by the accreting material. But how to transfer 
this power to a sizable fraction of ejected material remains a critical issue. It is now clear that the required 
driving power cannot be of thermal origin \citep{Matt07}. On the other hand, the push provided by turbulent Alfv\'en 
waves, such as those excited by the impact of the accretion streams onto the stellar surface, is likely to remain 
insufficient to drive massive stellar winds \citep{Cranmer08, Cranmer09}. But more importantly, it is quite tricky 
to assume that some accretion energy would be missing (the fraction that would possibly feed the stellar wind), 
while still explaining the observed UV luminosity. Indeed, it would imply an even higher mass flux onto the star, 
hence a higher spinning up torque \citep{ZanFer11}.

Another class of ejection phenomena is expected to arise because of the expansion and subsequent reconnection of 
the closed magnetospheric field lines. The inflation process is  the result of the star-disk differential rotation and the 
consequent build-up of toroidal magnetic field pressure. This is the same phenomenon that bounds the size of 
the magnetosphere connecting the star with the disk and limits the efficiency of the \citeauthor{Ghosh79} mechanism.
While semi-analytical models have foreseen the magnetic field expansion \citep[see e.g.][]{Aly90, UKL02}, different numerical 
experiments showed that plasma ejection is actually associated with the inflation process \citep{Hayashi96, Goodson97, 
Miller97, Romanova09}. Some observable properties of this phenomenon have been discussed, for example, by 
\citet{Hartmann09} and \citet{Gomez11}.  Besides, \citet{Hartmann02, Hartmann09} suggested that this mechanism could 
enhance the angular momentum loss from the star-disk system.

In this paper we present the results of a series of numerical MHD time-dependent simulations to analyze in detail
the energetics and dynamics of these magnetospheric ejections in different accretion regimes and evaluate their 
impact on the angular momentum balance of the star-disk system. At the same time, we are able to include in our models
the effects of stellar winds.
In Section \ref{sec:numerics} we present the numerical method and provide the initial and boundary conditions employed 
to carry out the numerical experiments. In Section \ref{sec:mej} we present in detail the outcome of a reference case: we first
characterize the dynamical properties of the simulated outflows (Section \ref{sec:mes} and \ref{sec:swdw}) 
and, subsequently, determine their influence on the angular momentum of the disk (Section \ref{sec:disktor}) 
and the star (Section \ref{sec:sangmom}). In Section \ref{sec:othercases} we study the impact of the  
disk accretion rate onto the dynamics of the outflows and the stellar spin evolution. In Section \ref{sec:discussion} we 
discuss the outcome of our models making a comparison with other scenarios proposed to solve the stellar spin 
conundrum. In Section \ref{sec:summary} we summarize our conclusions.

\section{Numerical setup}
\label{sec:numerics}

The models presented in this paper are numerical solutions of the magneto-hydrodynamic (MHD) system of equations, including
resistive and viscous effects:
\begin{eqnarray}
\label{eq:mhd}
\frac{\partial \rho}{\partial t} & + & \nabla\cdot(\rho \vec {u})  = 0 \nonumber \\
\frac{\partial \rho\vec{u}}{\partial t} & + &\nabla \cdot \left[ 
       \rho \vec{u}\vec{u} +
       \left( P + \frac{\vec{B}\cdot\vec{B}}{8\pi} \right)\vec{I}-
       \frac{\vec{B}\vec{B}}{4\pi} - \vec{\tau}
       \right]  = - \rho \nabla \Phi_\mathrm{g} \nonumber \\
\frac{\partial E}{\partial t} & + & \nabla\cdot\left[
    \left(E + P + \frac{\vec{B}\cdot\vec{B}}{8\pi}\right)\vec{u}-
    \frac{\left(\vec{u}\cdot\vec{B}\right)\vec{B}}{4\pi} \right] +  \\
   & + & \nabla\cdot\left[\eta_{\mathrm{m}}\vec{J}\times\vec{B}/4\pi -\vec{u}\cdot\vec{\tau} 
    \right] = - \rho \nabla \Phi_\mathrm{g} \cdot\vec{u}-\Lambda_\mathrm{cool} \nonumber \\
\frac{\partial \vec{B}}{\partial t} & + & \nabla \times \left(\vec{B}\times\vec{u} + \eta_{\mathrm{m}} \vec{J} \right)= 0 \nonumber \; .
\end{eqnarray}
This system expresses the conservation of mass, momentum and energy and includes the induction equation to
describe the evolution of the magnetic field. In the system of Eqs. (\ref{eq:mhd}) $\rho$ is the mass density, $\vec{u}$ is the flow speed, $P$ 
is the plasma thermal pressure, $\vec{B}$ is the magnetic field, $\Phi_\mathrm{g} = -GM_\star/R$ is the gravitational potential,
$\vec{J} = \nabla \times \vec{B}/4\pi$ is the electric current,  and $\eta_\mathrm{m}$ is the magnetic resistivity, where $\nu_\mathrm{m} = 
\eta_\mathrm{m}/4\pi$ defines the magnetic diffusivity.
The total energy density $E$ is defined as 
\[
E = \frac{P}{\gamma-1}+\rho\frac{\vec{u}\cdot\vec{u}}{2}+\frac{\vec{B}\cdot\vec{B}}{8\pi} \; ,
\]
where $\gamma = 5/3$ is the polytropic index of the plasma. The viscous stress tensor $\vec{\tau}$ is given by
\begin{equation}
\tau = \eta_\mathrm{v} \left[ \left(\nabla  \vec{u}\right) + \left(\nabla  \vec{u}\right)^\mathrm{T} - \frac{2}{3}\left(\nabla\cdot\vec{u}\right) \vec{I} \right] \; ,
\end{equation}
where $\eta_\mathrm{v}$ is the dynamic and $\nu_\mathrm{v} = \eta_\mathrm{v}/\rho$ is the kinematic viscosity. The anomalous transport
coefficients $\eta_\mathrm{m}$ and $\eta_\mathrm{v}$ are assumed to be of turbulent origin and parametrized according to an 
$\alpha$ prescription \citep{Shakura73}. The cooling term $\Lambda_\mathrm{cool} = \eta_\mathrm{m}\vec{J}\cdot\vec{J} + 
\mathrm{Tr}\left(\vec{\tau}\vec{\tau}^\mathrm{T}\right)/2\eta_\mathrm{v}$  is included to balance the viscous and Ohmic heating, so that the
system should evolve adiabatically, modulo numerical dissipative effects.
We employed the MHD module provided with the PLUTO code\footnote{PLUTO is freely downloadable at http://plutocode.ph.unito.it.} 
 \citep{Mignone07} to solve the system of Eqs. (1). For a precise description of the employed algorithm, we refer the reader to \citetalias{ZanFer09}.

\subsection{Initial and boundary conditions}

We employ the same initial and boundary conditions, computational domain and resolution of the simulations presented in \citetalias{ZanFer09},
where a more extensive discussion about the numerical setup can be found. For the sake of completeness, we recall here the main 
characteristics of our setup. 

The two-dimensional simulations are carried out in spherical coordinates ($R$, $\theta$) assuming axisymmetry around the rotation axis of the star.
We indicate the cylindrical radius $r = R\sin \theta$ and the height $z = R\cos \theta$ with lower-case letters.
We initially consider a viscous accretion $\alpha$-disk surrounded by a rarefied corona threaded by the stellar magnetosphere. The Keplerian 
accretion disk is modeled after the polytropic solution presented in \citet{KluzniakKita00} \citep[see also][]{RegevGit02,Umurhan06}. The density,
pressure, toroidal and accretion speed and kinematic viscosity of the disk are given respectively by:
\begin{eqnarray}
\rho_\mathrm{d} & = & \rho_\mathrm{d0} \left\{\frac{2}{5\epsilon^2}\left[\frac{R_\star}{R}-\left(1-\frac{5\epsilon^2}{2}\right)\frac{R_\star}{r}\right]\right\}^{3/2} \nonumber \\
P_\mathrm{d}  & = & \epsilon^2 \rho_\mathrm{d0} V^2_{\mathrm{K}\star} \left( \frac{\rho_\mathrm{d}}{\rho_\mathrm{d0}}  \right)^{5/3} \\
u_{\phi\mathrm{d}} & = & \left[ \sqrt{1-\frac{5\epsilon^2}{2}}+\frac{2}{3}\epsilon^2\alpha^2_\mathrm{v}\Lambda\left(1-\frac{6}{5\epsilon^2\tan^2\theta} \right) \right] \sqrt{\frac{GM_\star}{r}} \nonumber \\
u_{R\mathrm{d}} & = & -\alpha_\mathrm{v}\epsilon^2\left[10-\frac{32}{3}\Lambda\alpha_\mathrm{v}^2-\Lambda\left(5 -\frac{1}{\epsilon^2\tan^2\theta}\right)\right] \sqrt{\frac{GM_\star}{R\sin^3\theta}} \nonumber \\
\nu_\mathrm{v} & = & \frac{2}{3}\alpha_\mathrm{v} \left[ \left. C^2_\mathrm{s}\left(r\right) \right|_{z=0}+\frac{2}{5}\left( \frac{GM_\star}{R}-\frac{GM_\star}{r}\right)\right]\sqrt{\frac{r^3}{GM_\star}}  \; , \nonumber
\end{eqnarray}
where $\alpha_\mathrm{v}$ is the anomalous viscosity coefficient,  
$\Lambda = \left(5/11+64/55\,\alpha_\mathrm{v}^2\right)^{-1}$,
$C_\mathrm{s} = \sqrt{P/\rho}$ is the isothermal sound speed, 
$V_\mathrm{K} = \sqrt{GM_\star/r}$ is the Keplerian speed, $\epsilon = \left.C_\mathrm{s}/V_\mathrm{K}\right|_{z=0}$ is the disk aspect ratio, $\rho_\mathrm{d0}$ and $V_{\mathrm{K}\star}$ are the density and Keplerian 
speed on the midplane of the disk at $R=R_\star$. The magnetic diffusivity is assumed to be proportional to the kinematic viscosity:
\begin{equation}
\label{eq:res}
\nu_\mathrm{m} = \alpha_\mathrm{m} \frac{3}{2} \frac{\nu_\mathrm{v}}{\alpha_\mathrm{v}} \; ,
\end{equation}
so that the magnetic Prandtl number is equal to $\mathcal{P}_\mathrm{m} = \nu_\mathrm{v}/\nu_\mathrm{m} = 2\alpha_\mathrm{v}/3\alpha_\mathrm{m}$.

The corona is represented by a polytropic hydrostatic atmosphere whose density and pressure distributions are given by:
\[
\rho_\mathrm{a} = \rho_{\mathrm{a}0}\left( \frac{R_\star}{R} \right)^{3/2} \qquad 
P_\mathrm{a} = \frac{2}{5}\rho_{\mathrm{a}0}\frac{GM_\star}{R_\star}\left( \frac{R_\star}{R} \right)^{5/2} \; ,
\]
where $\rho_{\mathrm{a}0} \ll \rho_{\mathrm{d}0}$ is the density of the corona on the spherical surface $R=R_\star$.

We model the stellar magnetosphere as a purely dipolar field aligned with the stellar rotation axis. Given the flux function $\Psi_\star$:
\begin{equation}
\Psi_\star = B_\star R_\star^3 \frac{\sin^2\theta}{R} \; ,
\end{equation}
the field components are defined as:
\[
B_R = \frac{1}{R^2 \sin\theta} \frac{\partial\Psi_\star}{\partial\theta} \qquad B_\theta =  -\frac{1}{R \sin\theta} \frac{\partial\Psi_\star}{\partial R} \; ,
\]
where $R_\star$ and $B_\star$ are the stellar radius and the magnetic field intensity at the stellar equator respectively.  The magnetic flux 
through one stellar hemisphere is equal to:
\begin{eqnarray}
\label{eq:sflux}
\Phi_\star & = & 2 \pi R_\star^2 \int_0^{\pi/2}  B_R\left(R_\star ,\theta \right)  \sin\theta \, \mathrm{d}\theta  \\
                 & = & 2\pi \left[ \Psi_\star\left(R_\star,\pi/2\right) - \Psi_\star\left(R_\star,0\right) \right] \; = \; 2\pi B_\star R_\star^2 \nonumber 
\end{eqnarray}

The disk surface is determined by the pressure equilibrium $P_\mathrm{d} = P_\mathrm{a}$, while the disk is initially truncated where 
$B^2/8\pi = P_\mathrm{d}$. In order to minimize initial transient effects due to the differential rotation between the disk and the corona
we set the magnetic surfaces anchored inside the Keplerian disk to corotate with it. The coronal density is corrected so as to nullify the
centrifugal acceleration perpendicularly to the magnetic surfaces. 

The computational domain encompasses a spherical sector going from the polar axis ($\theta = 0$) to the disk midplane ($\theta = \pi/2$) and
from an inner radius $R=R_\star$ up to $R =  28.6 R_\star$. The domain is discretized with a grid of $N_\theta \times N_R =  100 \times 214$
cells. The grid is stretched in the radial direction so that the cell sizes satisfy the condition $\Delta R \sim R\Delta \theta$. Suitable boundary
conditions are imposed to satisfy the axial and equatorial symmetries.
The boundary conditions on the stellar surface $R = R_\star$ are carefully chosen to model a perfect conductor rotating with an angular speed 
$\Omega_\star$ so that in the rotating frame of reference the electric field $\left.\vec{E}\right|_{\Omega = \Omega_\star} = \vec{B}\times\left(\vec{u} - 
\vec{\Omega_\star}\times\vec{R}  \right) = 0$ is zero. Besides, the boundary is designed in order to absorb the accretion funnels while forcing
the rarefied plasma of the surrounding corona to have a density and enthalpy suitable to drive a light stellar wind. At the outer boundary the 
variables are extrapolated, also ensuring that the area of this boundary directly connected to the central star exerts no artificial torques on the latter. 
A detailed description and discussion of the boundary conditions is given in \citetalias{ZanFer09}.

\subsection{Units and normalization}
\label{sec:units}

We performed the simulations and we are going to present their outcome in dimensionless units. We here provide the normalization factors
necessary to express the results in physical units, taking into account the typical case of a young forming star. The stellar radius $R_\star$ is
employed as unit length while, given the stellar mass and radius, the velocities are expressed in units of the Keplerian speed at the stellar 
surface $V_{\mathrm{K}\star} = \sqrt{GM_\star/R_\star}$. Taking $\rho_\mathrm{d,0}$ as the normalization density, the magnetic field is given
in units of $\sqrt{\rho_\mathrm{d0} V_{\mathrm{K}\star}^2}$, time in units of $t_0 = R_\star/V_{\mathrm{K}\star}$, accretion and ejection rates in
units of $\dot{M}_0 = \rho_{\mathrm{d}0}R_\star^2V_{\mathrm{K}\star}$, powers in units of 
$\dot{E}_0 = \rho_{\mathrm{d}0}R_\star^2V_{\mathrm{K} \star}^3$ and torques in units of $\dot{J}_0 = \rho_{\mathrm{d}0}R_\star^3V_{\mathrm{K} \star}^2$.
Assuming $M_\star = 0.5 M_\odot$,  $R_\star = 2R_\odot$ and $\rho_{\mathrm{d}0} = 8.5\times10^{-11}$g cm$^{-3}$, we have that:
\begin{eqnarray}
V_{\mathrm{K} \star} & = & 218 \left(\frac{M}{0.5 M_\odot}\right)^{1/2} \left( \frac{R_\star}{2 R_\odot} \right)^{-1/2}\; \mathrm{\mbox{km s}}^{-1}  \nonumber \\
B_0 & = & 200 \left( \frac{\rho_{\mathrm{d}0}}{\scriptstyle 8.5\times 10^{-11} \; \mathrm{\scriptsize \mbox{g cm}}^{-3}} \right)^{1/2} \left(\frac{M}{0.5 M_\odot}\right)^{1/2} \left( \frac{R_\star}{2 R_\odot}\right)^{-1/2} \; \mathrm{\mbox{G}} \nonumber \\
t_0 & = & 0.074 \left(\frac{M}{0.5 M_\odot}\right)^{-1/2}  \left( \frac{R_\star}{2 R_\odot} \right)^{3/2} \; \mathrm{\mbox{days}} \\
\dot{M}_0 & = & 5.7 \times 10^{-7} \left(\frac{\rho_{\mathrm{d}0}}{\scriptscriptstyle 8.5\times10^{-11} \; \mathrm{\mbox{\scriptsize g cm}}^{-3}}\right) 
\left(\frac{M}{0.5M_\odot}\right)^{1/2} \left( \frac{R_\star}{2R_\odot} \right)^{3/2} 
                     \; M_\odot \; \mathrm{\mbox{yr}}^{-1} \nonumber
\end{eqnarray}
In order to directly provide the characteristic spin-up/spin-down timescales, the torques acting onto the star will be divided by the stellar angular momentum,
expressed in units of $J_{\star 0} = M_\star R_\star V_{\mathrm{K}\star}$. The inverse of the characteristic braking timescale will therefore be expressed in units of
\footnote{This normalization differs from the one employed in \citetalias{ZanFer09}, where we normalized the stellar angular momentum in units of 
$J_{\star 0} =  \rho_{\mathrm{d}0} R_\star^4 V_{\mathrm{K}\star}$, so that the braking timescale would be given in units of $t_0$.}:
\begin{equation}
\left.\frac{\dot{J}}{J_\star}\right|_0 = 10^{-6}\,  \left( \frac{\rho_{\mathrm{d}0}}{\scriptstyle 8.5\times 10^{-11} \; \mathrm{\scriptsize \mbox{g cm}}^{-3}} \right)
\left(\frac{M}{0.5 M_\odot}\right)^{-1/2} \left( \frac{R_\star}{2 R_\odot} \right)^{3/2} \; \mathrm{yr}^{-1}
\label{eq:tnormdens}
\end{equation}

\begin{table}[t]
\caption{Parameters of the simulations: viscosity coefficient $\alpha_\mathrm{v}$, magnetic resistivity coefficient $\alpha_\mathrm{m}$, magnetic Prandtl number
                $\mathcal{P}_\mathrm{m} = \eta_\mathrm{v}/\eta_\mathrm{m}$, initial viscous accretion rate of the Keplerian disk $\dot{M}_\mathrm{d}$.} 
\label{table:cases}
\centering                   
\begin{tabular}{c c c c c}   
\hline\hline        
Simulation & $\alpha_\mathrm{v}$ & $\alpha_\mathrm{m}$ & $\mathcal{P}_\mathrm{m}$ &$ \dot{M}_\mathrm{d}/\dot{M}_0$ \\   
\hline                  
C1 &  1 & 0.1 & 6.7 &$1.4\times 10^{-2}$ \\
C03 &  0.3 & 0.1 & 2 &$4.2\times 10^{-3}$ \\
C01 &  0.1 & 0.1 & 0.67 &$1.4\times 10^{-3}$\\
E1 &   1   & 1 & 0.67 &$1.4\times 10^{-2}$ \\ 
\hline                                   
\end{tabular}
\end{table}

\subsection{The simulations}

Once the initial conditions are normalized, the problem depends on six dimensionless parameters: the disk thermal scale height $\epsilon$, the equatorial stellar field
intensity $B_\star/B_0$, the stellar rotation rate $\delta_\star = R_\star \Omega_\star/V_{\mathrm{K} \star}$, the coronal density contrast $\rho_{\mathrm{a}0}/\rho_{\mathrm{d}0}$,
the viscous and resistive coefficients $\alpha_\mathrm{v}$ and $\alpha_\mathrm{m}$. Except for the transport coefficients, the other parameters are the same 
used in the simulations of \citetalias{ZanFer09}: $\epsilon = 0.1$, $B_\star = 5 B_0$, $\delta_\star = 0.1$ and $\rho_{\mathrm{a}0}/\rho_{\mathrm{d}0} = 10^{-2}$.
Using the standard normalization given in Section \ref{sec:units}, this corresponds to a stellar magnetic field $B_\star = 1$ kG and a period of rotation of the
star $P_\star = 2\pi t_0/\delta_\star = 4.65$ days with a Keplerian corotation radius $R_\mathrm{co} = R_\star/\delta_\star^{2/3} = 4.64 \, R_\star$. 

The transport coefficients $\alpha_\mathrm{v}$ and $\alpha_\mathrm{m}$ control, respectively, the intensity of the viscous torque allowing the disk to accrete and the
strength of the coupling of the stellar magnetic field with the disk material. As discussed, for example, in \citet{UKL02} and \citet{Matt05a}, the disk magnetic resistivity 
controls the extent of the disk region that is steadily connected to the star: since the opening of the magnetosphere is determined by the star-disk differential rotation
and the consequent buildup of toroidal magnetic pressure, a weaker magnetic coupling (i.e. a higher $\alpha_\mathrm{m}$) limits the growth of the toroidal field and
therefore increases the size of the connected region. For example, in \citetalias{ZanFer09} we had to assume a value $\alpha_\mathrm{m} = 1$ in order to maintain the magnetic 
connection beyond the corotation radius (``extended'' magnetosphere). 

Since the main aim of this paper is to study the dynamical processes associated with the inflation and opening of the magnetospheric field lines, we assume a stronger 
magnetic coupling ($\alpha_\mathrm{m} = 0.1$), so that the magnetic configuration opens up closer to the stellar surface (``compact'' magnetosphere), 
where these phenomena can strongly affect both the disk and the stellar dynamics. Besides, we consider different values of the viscosity coefficient 
$\alpha_\mathrm{v}$ = 1, 0.3, 0.1 in order to study the evolution of star-disk systems characterized by different accretion rates. 
The summary of the cases presented in this paper is given in Table \ref{table:cases}. We also include
the parameters characterizing the reference case of \citetalias{ZanFer09} (case E1) that will be considered to do some comparisons.
   
  \begin{figure}[!t]
  \resizebox{\hsize}{!}{\includegraphics*{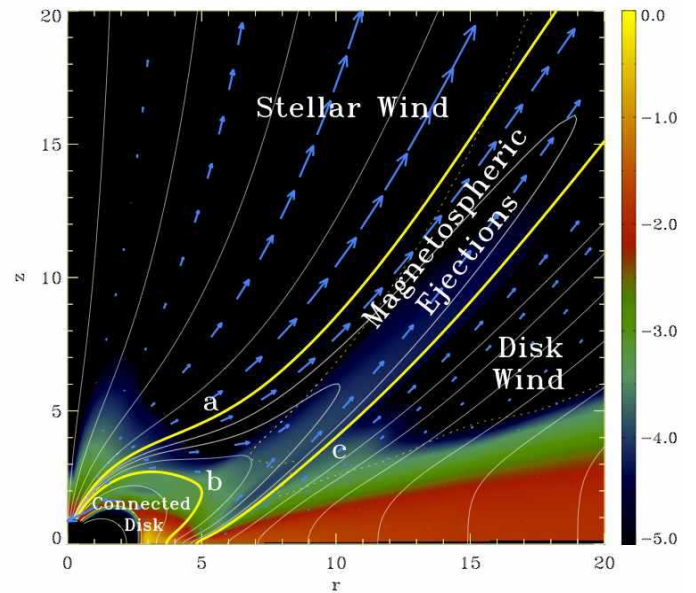}}
  \caption{Global view of the star-disk interacting system. A logarithmic density map is shown in the background. Poloidal speed vectors are 
                  represented as blue arrows. The dotted line marks the Alfv\'en surface, where $u_\mathrm{p} = B_\mathrm{p}/\sqrt{4\pi \rho}$.
                  Sample field lines are plotted with white solid lines. Thick yellow field lines, labeled as (a), (b), and (c), delimit the different dynamical
                  consituents of the system indicated in the figure. The image has been obtained by averaging in time the simulation outcome over 54
                  stellar periods.}
  \label{fig:globalview}
  \end{figure}
         
\section{Star-disk interaction and magnetospheric ejections}
\label{sec:mej}
 
  \begin{figure*}[!t]
  \centering
  \includegraphics[width=17cm]{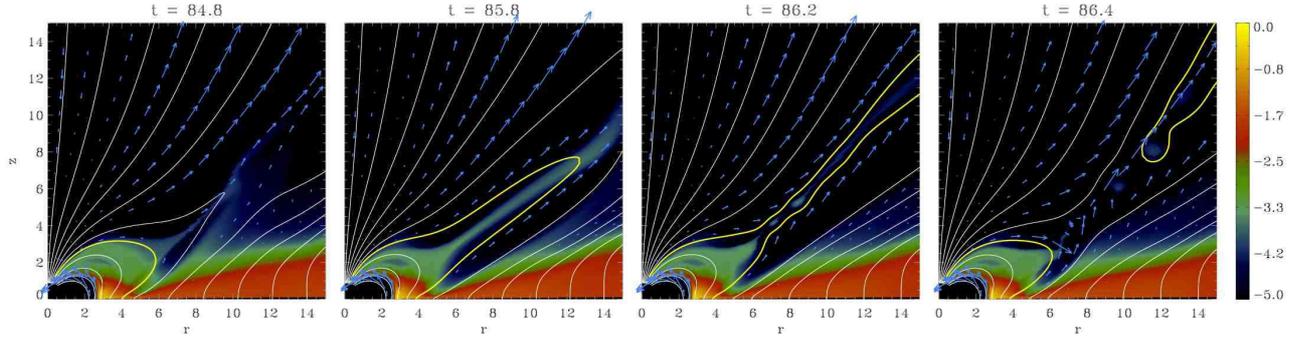}
   \caption{Temporal evolution of the periodic inflation/reconnection process which characterizes the dynamics of the magnetospheric
                   ejections in case C03. We show logarithmic density maps with sample field lines (white solid lines) and poloidal speed vectors (blue arrows)
                   superimposed. The yellow solid lines follow the evolution of a single magnetic surface showing clearly the dynamics of the
                   phenomenon. Time is given in units of rotation periods of the central star.}
   \label{fig:Alpha03cycle}
  \end{figure*}
   
Using case C03 as a representative example, we are going to analyze the dynamical properties of a interacting star-disk system 
in which the stellar magnetic field is strongly coupled to the accretion disk. As already pointed out, in such a situation the star-disk 
differential rotation generates a strong toroidal field component and, due to its pressure, the magnetic structure relaxes 
by inflating and opening the initial dipolar configuration close to the truncation radius.
The overall picture illustrating the outcome of this process is given in Fig. \ref{fig:globalview}. Four groups of field lines can be 
distinguished: (1) the field lines steadily connecting the disk with the star below the magnetic surface (b); (2) the open field 
lines anchored on the surface of star at latitudes higher than the position of the surface (a); (3) the open field lines attached to 
the accretion disk beyond the surface (c);  (4) the field lines enclosed between surfaces (a), (b) and (c) connecting the disk with 
the star, periodically evolving through stages of inflation, reconnection and contraction. An example of this periodic process 
is represented in Fig. \ref{fig:Alpha03cycle}. The periodicity of these phenomena corresponds to about 2 stellar rotation periods. 
Anyway, since the reconnection processes which are involved are driven by numerical resistivity, this periodicity has to be 
considered cautiously. On the other hand, this almost periodical behavior allows us to use time averages to characterize 
the long term evolution of the system and smooth out transient features. For example, Fig. \ref{fig:globalview} has been 
obtained by averaging snapshots over 54 rotation periods of the star: notice that, because of the time average, the fleeting 
reconnection phenomena are not visible. 
  
Different dynamical processes are associated with the four groups of field lines. In the region inside the magnetic surface (b),  
(``connected disk'') the star and the disk can directly exchange angular momentum, the disk is truncated and the accretion curtains 
form. Notice that this region extends within the Keplerian corotation radius $R_\mathrm{co} = 4.64 R_\star$ so that, beyond this radius, the disk and the star do 
not have a direct magnetic connection. Therefore, the \citeauthor{Ghosh79} scenario cannot be directly applied:  the disk region 
beyond corotation, which rotates slower than the star, cannot exert any direct spin-down torque onto the star. Three classes of 
outflows correspond to the other groups of field lines. A stellar wind flows along the open magnetic surfaces anchored at high 
stellar latitudes. A disk wind can be accelerated along the disk open field lines, but only the field lines closer to the star are 
characterized by a field strong enough for this outflowing component to play a relevant role, as it will be shown in Section \ref{sec:disktor}.
Finally, the inflation at mid-latitudes of the dipolar field lines is very dynamic and it is accompanied by outflows that can in principle 
extract mass, energy and angular momentum  both from the disk and the star. On a relatively large scale (10-20 $R_\star$, see 
Fig. \ref{fig:Alpha03cycle}), these ejections detach from the magnetosphere in a reconnection event and continue their propagation 
as magnetic islands disconnected from the central part of the star-disk system, in between the open magnetic surfaces anchored 
into the star and those anchored into the disk. In the following we will refer to this type of outflow associated with the process of 
inflation/reconnection of the magnetospheric field lines as \emph{magnetospheric ejections} (MEs).

\begin{table}[b]
\caption{Fraction of stellar poloidal magnetic fluxes threading different regions of the star-disk system in different cases:  
                stellar wind $\Phi_\mathrm{SW}$, MEs $\Phi_\mathrm{ME}$, steadily connected disk $\Phi_\mathrm{CD}$,
                magnetospheric cavity $\Phi_\mathrm{MC}$, accretion columns $\Phi_\mathrm{AC}$. Also shown are
                the fractional surface of the accretion columns $S_\mathrm{AC}$, the position of the truncation radius $R_\mathrm{t}$,
                the anchoring radii at the disk midplane of the outermost closed ($R_\mathrm{cm}$) and the innermost open 
                ($R_\mathrm{om}$) magnetic surfaces. Different estimates have been done for the accretion and propeller 
                phases of case C01 (see Section \ref{sec:propeller}).}
\label{table:fluxes}
\centering
\begin{tabular}{c c c c c c} 
\hline\hline
 & C1 & C03 & C01 (acc.) & C01 (pro.) & E1\\ 
 \hline 
$\Phi_\mathrm{SW}/\Phi_\star$ & 0.052 & 0.066  & 0.091 & 0.109 & 0.091 \\
$\Phi_\mathrm{ME}/\Phi_\star$ & 0.023 & 0.035 & 0.039 & 0.053 &  0.009 \\
$\Phi_\mathrm{CD}/\Phi_\star$  & 0.269 & 0.180  & 0.135 & 0.102 & 0.239 \\           
$\Phi_\mathrm{MC}/\Phi_\star$  & 0.656 & 0.719  & 0.735 & 0.736 & 0.661 \\
\hline
$\Phi_\mathrm{AC}/\Phi_\star$  & 0.194 & 0.135  & 0.087 & 0. & 0.152 \\
$S_\mathrm{AC}/S_\star$           & 0.110 & 0.075 & 0.049 &  0. & 0.088 \\
\hline
$R_\mathrm{t}/R_\star$ & 1.9 & 2.7 & 3.0 & 4.4 & 2.5 \\
$R_\mathrm{cm}/R_\star$  & 2.7 & 3.7 & 3.8 & 4.7 & 11.9 \\
$R_\mathrm{om}/R_\star$ & 3.1 & 4.7 & 4.2 & 5.7 & 15.1 \\
\hline 
\end{tabular}
\end{table}

The different regions outlined in Fig. \ref{fig:globalview} can be also characterized by the amount of poloidal magnetic flux which 
participates in each of them. In Table \ref{table:fluxes} we show for all the discussed cases the fraction of magnetic flux 
$\Phi = \int \vec{B}_\mathrm{p} \cdot \mathrm{d} \vec{S}$ which crosses each region relative to the total stellar flux through one 
hemisphere, Eq. (\ref{eq:sflux}): we display the stellar wind flux $\Phi_\mathrm{SW}$, also equal to the open 
magnetic flux of the disk wind, the MEs flux $\Phi_\mathrm{ME}$, the flux of the connected disk $\Phi_\mathrm{CD}$ and the 
flux contained in the magnetic cavity inside the truncation radius $\Phi_\mathrm{MC}$. Obviously we have $\Phi_\mathrm{SW}+
\Phi_\mathrm{ME}+\Phi_\mathrm{CD}+\Phi_\mathrm{MC} = \Phi_\star$. Since not all the field lines steadily connecting the star to
the disk are mass-loaded to form the accretion funnels, we also show the amount of magnetic flux threading the accretion columns
$\Phi_\mathrm{AC}$, clearly equal to a fraction of $\Phi_\mathrm{CD}$, and the corresponding surface covering fraction 
$S_\mathrm{AC}/S_\star$. To give an indication of the size of the different interaction regions we also provide the position of the disk
truncation radius $R_\mathrm{t}$,  the anchoring radius $R_\mathrm{cm}$ of the outermost steadily connected magnetic surface 
(labeled as (b) in  Fig. \ref{fig:globalview}) and the anchoring radius $R_\mathrm{om}$ of the innermost open magnetic surface 
(labeled as (c) in Fig. \ref{fig:globalview}) at the disk midplane.  Notice that, while the magnetic flux is frozen into the stellar
surface, its distribution can change on the disk midplane. Besides providing a clear indication about the relative
importance of the different dynamical components, these quantities can be directly compared to the predictions of other models, such as 
the X-Wind.

\subsection{Dynamical properties of magnetospheric ejections}
\label{sec:mes}

We are going to characterize the properties of the MEs by inspecting their mass, angular momentum and energy fluxes.
By defining a surface $\vec{S}$ perpendicular to the poloidal flow these are respectively defined as:
\begin{eqnarray}
\dot{M} & = & \int_{\vec{S}} \rho \vec{u}_\mathrm{p} \cdot \mathrm{d} \vec{S} \label{eq:mflux} \\ 
\dot{J} & = & \int_{\vec{S}} \left( r\rho u_\phi \vec{u}_\mathrm{p} - \frac{rB_\phi\vec{B}_\mathrm{p}}{4 \pi} \right) \cdot  \mathrm{d} \vec{S} \label{eq:jflux} \\ 
\dot{E} & = & \int_{\vec{S}} \left[ \left(\frac{1}{2} \rho u^2 + \frac{\gamma P}{\gamma-1}-\frac{GM_\star}{R}\right)\vec{u}_\mathrm{p}   +  \vec{E}\times\vec{B}\big|_\mathrm{p} \right] \cdot \mathrm{d} \vec{S} \label{eq:eflux}
\end{eqnarray}
As already pointed out, this type of ejection can exchange mass, angular momentum and energy both with the star and the
disk. We can therefore define different contributions to the budget of the MEs. Taking for example the mass flux, we can estimate
the stellar mass input as:
\begin{equation}
\dot{M}_\mathrm{ME,s} = 4\pi R_\star^2 \int_{\theta_\mathrm{a}}^{\theta_\mathrm{b}} \rho u_R \sin\theta \, \mathrm{d}\theta \; ,
\label{eq:MEstar}
\end{equation}
where $\theta_\mathrm{a}$ and $\theta_\mathrm{b}$ are the anchoring angles onto the stellar surface of the lowest open and the outermost steadily 
closed magnetic surfaces respectively, labeled as (a) and (b) in Fig. \ref{fig:globalview}. The disk contribution can be calculated as:
\begin{equation}
\dot{M}_\mathrm{ME,d} = -4 \pi \int_{R_\mathrm{in}}^{R_\mathrm{out}} \rho^+u_p^+ \; r\mathrm{d}r \; ,
\end{equation}
where $R_\mathrm{in}$ and $R_\mathrm{out}$ are the anchoring radii at the disk surface of the outermost steadily closed magnetic surface and the innermost
open field line threading the disk, labeled as (b) and (c) in Fig. \ref{fig:globalview}, whose footpoint radii at the disk midplane are $R_\mathrm{cm}$ and 
$R_\mathrm{om}$ respectively (see Table \ref{table:fluxes}). These radii are marked in the panels of Fig. \ref{fig:Comp03} and \ref{fig:Comp1}.
The $+$ superscript indicates a quantity evaluated at the disk surface 
$H\left(r\right) = 2 C_\mathrm{s}/\Omega_\mathrm{K}$, proportional to the thermal heightscale;  the poloidal vectors labeled with a $+$ are defined as the 
components perpendicular to the surface of the disk: for example, in the case of the speed, 
$u_\mathrm{p}^+ = \left. -u_z + u_r H'  \right|_{H\left( r \right)} = \left. u_\theta\left( \sin\theta+\cos\theta H' \right) - u_R  \left( \cos\theta - \sin \theta H' \right)  \right|_{H\left( r \right)}$.
The total mass flux of the MEs can be evaluated by choosing a surface crossing both magnetic field lines (a) and (c); for example, by selecting
a spherical zone with a radius $R \geq 6R_\star$, so that it crosses both magnetic surfaces (a) and (c), we define:
\begin{equation}
\dot{M}_\mathrm{ME,tot} = 4\pi R^2 \int_{\theta_\mathrm{a}}^{\theta_\mathrm{c}} \rho u_R \sin\theta \, \mathrm{d}\theta \; ,
\label{eq:MEtot}
\end{equation}
where $\theta_\mathrm{a}$ and $\theta_\mathrm{c}$ are the angles at which magnetic surfaces (a) and (c) intercept the sphere with radius $R$. 
When calculating the time evolution of the previous fluxes, the anchoring angles and radii can vary with time.
In a steady situation or taking into account time-averaged quantities for a system that evolves periodically (as in the considered case), we have that
$\dot{M}_\mathrm{ME,tot} = \dot{M}_\mathrm{ME,s} + \dot{M}_\mathrm{ME,d}$.
Analogous expressions can be derived for the angular momentum ($\dot{J}_\mathrm{ME,s}, \dot{J}_\mathrm{ME,d}, \dot{J}_\mathrm{ME,tot}$) and energy
fluxes ($\dot{E}_\mathrm{ME,s}, \dot{E}_\mathrm{ME,d}, \dot{E}_\mathrm{ME,tot}$) by integrating Eq. (\ref{eq:jflux}) and (\ref{eq:eflux}) on the same
surface elements employed to define the mass fluxes. Notice that we exploited the midplane symmetry of our simulations so that the flux integrals
are related to two-sided outflows.

 \begin{figure}[!t]
 \resizebox{\hsize}{!}{\includegraphics*{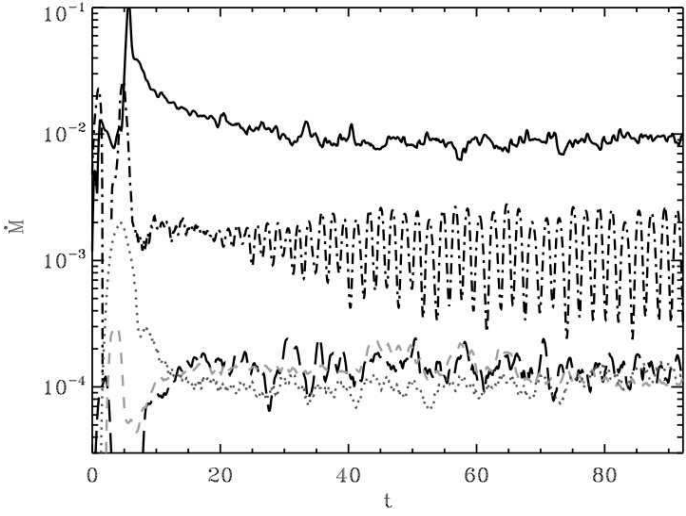}}
 \caption{Temporal evolution of mass fluxes of the different accretion and ejection phenomena present in the system: mass accretion rate measured at
                 the stellar surface ({\it solid line}), total mass outflow rate of magnetospheric ejections ({\it dot-dashed line}), mass flux fueling the MEs coming
                 from the star only ({\it long-dashed line}), stellar wind outflow rate ({\it dashed grey line}), disk wind mass outflow rate ({\it dotted line}).
                 The image refers to case C03. Time is given in units of rotation periods of the central star.}
 \label{fig:Alpha03rates}
 \end{figure}

\subsubsection{Mass fluxes}
\label{sec:mflux} 

In Fig. \ref{fig:Alpha03rates} we plot the temporal evolution of the mass-loss rates of the different outflowing components and compare
them to the mass accretion rate measured onto the surface of the star, defined as:
\begin{equation}
\dot{M}_\mathrm{acc,s} = -4\pi R_\star^2 \int_{\theta_\mathrm{b}}^{\pi/2} \rho u_R \sin\theta \, \mathrm{d}\theta \; .
\label{eq:maccs}
\end{equation}
After an initial transient lasting around 20 stellar rotation periods, the total outflow rate of the MEs, calculated using Eq.  (\ref{eq:MEtot}) with $R=7R_\star$,
regularly oscillates around a value corresponding to $\approx 18\%$ of the accretion rate at the stellar surface. This corresponds to 
$\approx 15\%$ of the disk accretion rate measured at $R_\mathrm{out}$. The oscillations correspond to the periodic inflation 
and reconnection phenomena. The stellar contribution to the MEs, Eq. (\ref{eq:MEstar}), is of the order of a few percent of the accretion rate.
This means that MEs are essentially mass-loaded from the disk and their inertia is dominated by material coming from the accretion disk.
The stellar and disk wind mass fluxes plotted in Fig. \ref{fig:Alpha03rates} will be defined and discussed in Section \ref{sec:swdw}.

 \begin{figure}[!t]
 \resizebox{\hsize}{!}{\includegraphics*{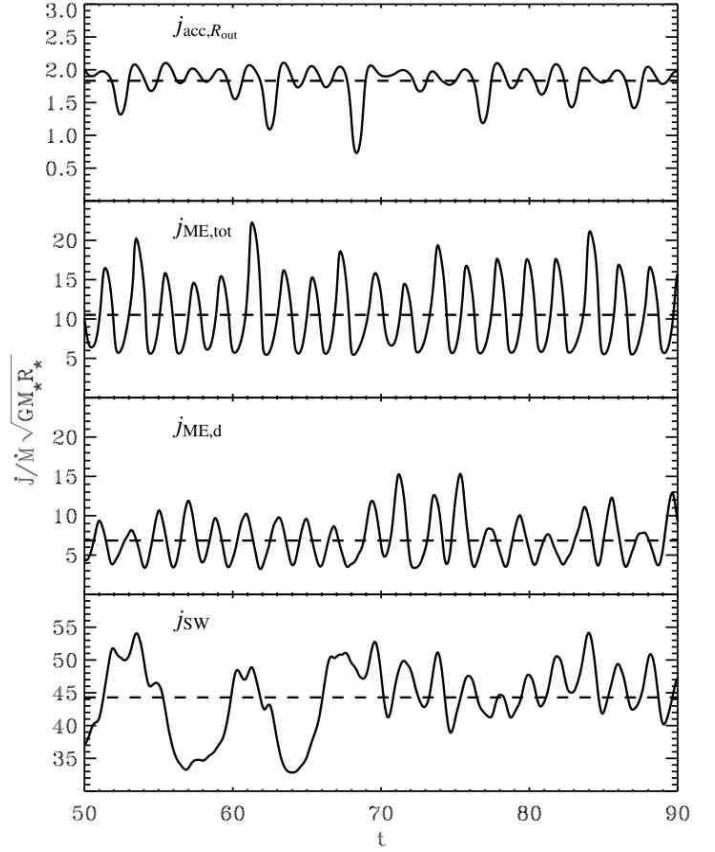}}
 \caption{Temporal evolution of the specific angular momentum carried by different accretion and ejection component of the system in case C03: specific angular 
                 transported by the accretion flow through $R_\mathrm{out}$ (upper panel); total specific angular momentum carried by the MEs (second panel);
                 specific angular momentum extracted by the MEs from the disk only (third panel); specific angular momentum of the stellar wind (lower panel).
                 Dashed lines show the temporal averages of the quantities over the plotted lapse of time. Time is given in units of rotation periods of the star.}
 \label{fig:Alpha03lever}
 \end{figure}
 
\subsubsection{Angular momentum fluxes}
 
Since the MEs are magnetically connected both to the star and the disk, they can potentially extract angular 
momentum from both. In the second panel from the top in Fig. \ref{fig:Alpha03lever} we plot the temporal evolution of the 
normalized total specific angular momentum carried by the MEs, defined as:
\begin{equation}
j_\mathrm{ME,tot} = \frac{\dot{J}_\mathrm{ME,tot}}{\dot{M}_\mathrm{ME,tot}} \approx \frac{\dot{J}_\mathrm{ME,tot}}{\dot{M}_\mathrm{ME,d}} \; ,
\label{eq:jmet}
\end{equation}   
while in the third panel from the top we have evaluated the specific angular momentum extracted by the MEs from the disk only:
\begin{equation}
j_\mathrm{ME,d} = \frac{\dot{J}_\mathrm{ME,d}}{\dot{M}_\mathrm{ME,d}} \; .
\label{eq:jmed}
\end{equation}
Plots are in units of $\sqrt{GM_\star R_\star}$.
Both quantities regularly oscillates in time but, if we look at the time averages, we clearly see that
the specific angular momentum extracted from the disk is around $j_\mathrm{ME,d} \approx 7 \sqrt{GM_\star R_\star}$, while the total angular 
momentum of the MEs is approximately $j_\mathrm{ME,tot} \approx 10.5 \sqrt{GM_\star R_\star}$. Clearly, the total angular momentum of the 
MEs is larger than the angular momentum extracted from the disk only, meaning that a substantial fraction comes also from the star. 
In Section \ref{sec:disktor} and \ref{sec:sangmom} we will discuss the effects of these torques on the angular momentum evolution of the disk and the star respectively. 

 \begin{figure}[!t]
 \resizebox{\hsize}{!}{\includegraphics*{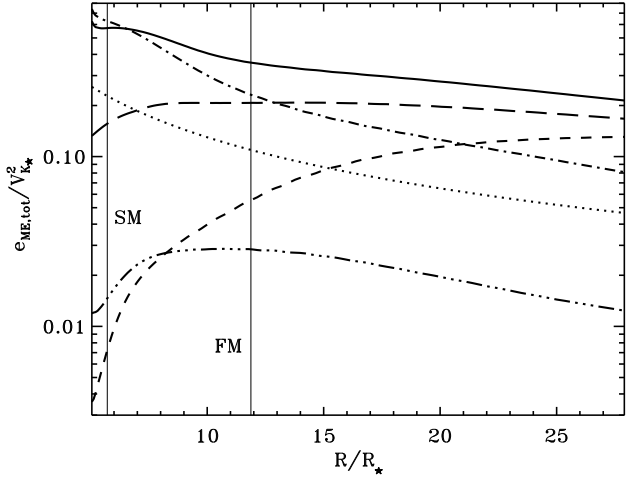}}
 \caption{Radial evolution of the total specific energy carried by the MEs in case C03. The total specific energy ({\it solid line}) is given by the sum of
                 Poyinting-to-mass flux ratio ({\it dot-dashed line}), kinetic energy ({\it long-dashed line}), specific enthalpy ({\it triple-dotted-dashed line})
                 and potential gravitational energy ({\it dotted line}, plotted in absolute value). The specific poloidal kinetic energy is also plotted
                 ({\it dashed line}). The slow- and fast-magnetosonic surfaces are marked by a vertical line. The plot starts from the cusp of the innermost
                 magnetic surface that steadily connnects the disk and the star (labeled as (b) in Fig. \ref{fig:globalview}). The figure has been obtained
                 by time averaging the energy and mass fluxes over 54 rotation periods of the star, from time $t =38$ up to $t=92$.}
 \label{fig:Invarmej}
 \end{figure}
 
\subsubsection{Energy fluxes}
   
Concerning the energy budget of the MEs, the power extracted from the disk only corresponds to $\dot{E}_\mathrm{ME,d} \approx 0.2 GM_\star\dot{M}_\mathrm{acc,s}/R_\mathrm{in}
\approx 0.05 GM_\star\dot{M}_\mathrm{acc,s}/R_\star$. It is important to point out that the mechanical power released by the material accreting from $R_\mathrm{out}$ 
down to $R_\mathrm{in}$, defined as
\begin{equation}
\dot{E}_\mathrm{acc} \approx \dot{M}_\mathrm{acc} \left. \left( \frac{u^2}{2} + \Phi_\mathrm{g} \right)\right|_{R_\mathrm{out}}-
                                                      \dot{M}_\mathrm{acc} \left. \left( \frac{u^2}{2} + \Phi_\mathrm{g} \right)\right|_{R_\mathrm{in}}
\label{eq:genber}
\end{equation}
is sufficient to power the part of the MEs coming from the disk. The power extracted from the star is comparable and largely determined by the Poynting 
flux associated with the spin-down torque $\dot{J}_\mathrm{ME,s}$. The enthalpy flux, needed in our simulations to give the initial drive to any stellar 
outflow, is a small fraction ($\approx 0.01 GM_\star\dot{M}_\mathrm{acc,s}/R_\star$) of the accretion power.
In case C03, the MEs operate as a magnetic sling, powered both by the stellar and disk rotation.

In order to have a better understanding of the asymptotic properties and the acceleration efficiency of MEs, we can inspect 
the energy conversion along the flow. In Fig. \ref{fig:Invarmej} we plot as a function of the radial coordinate $R$  the evolution along the flow of the total specific energy of the MEs:
\begin{equation}
e_\mathrm{ME,tot} = \frac{\dot{E}_\mathrm{ME,tot}}{\dot{M}_\mathrm{ME,tot}} \; .
\end{equation}
This plot has been obtained by averaging the mass and energy fluxes over 54 stellar periods and it starts from the cusp of field line (b), located at $\approx 5 R_\star$.
In a stationary situation this definition corresponds obviously to an average of the Bernoulli invariant over a section of the outflow;
\begin{equation}
e = \frac{1}{2}u^2+h-\frac{GM_\star}{R}-\frac{r\Omega_\star B_\phi B_p}{4\pi \rho u_p} \; ,
\label{eq:invaren}
\end{equation}
given by the sum of kinetic energy, enthalpy $h = \gamma P/(\gamma-1) \rho$, gravitational and magnetic (Poynting) energy.
On the other hand, the more general definition Eq. (\ref{eq:genber}) allow us to define an energy conversion efficiency in the case of non-stationary MEs. 
We can clearly see that the total energy is not conserved along the flow: this is mainly due to to the dissipation
of the magnetic energy due to the reconnection events. Even if the reconnection is controlled by numerical dissipation and therefore is not physical, it leads to 
a temperature increase, visible at lower radii in Fig. \ref{fig:Invarmej}. The outflow cools down subsequently: since we did not include any 
realistic cooling function, we simply limited the maximum specific entropy ($P/\rho^\gamma$) that can be attained by the outflow in order to preserve code stability.
Notice that the kinetic energy stops increasing after $\sim 10 R_\star$: this reflects the fact that, after the plasmoids have detached from the
inner magnetosphere, the acceleration process stops and the propagation of the outflow becomes ballistic. These aforementioned effects limit the acceleration
efficiency of the outflow, whose terminal speed is around 0.5 $V_{\mathrm{K}\star}$. 
Finally, consistently with their ballistic propagation, MEs cannot be self-collimated thanks to magnetic stresses, as usually envisaged for magnetically driven outflows. This
confers the characteristic ``conical'' shape to these magnetospheric outflows, as already pointed out by \citet{Romanova09}. On the other hand, they can be in principle 
confined by some external agent, as it will be discussed in Appendix \ref{sec:current}.

  \begin{figure*}[!t]
   \includegraphics[width=0.5\textwidth]{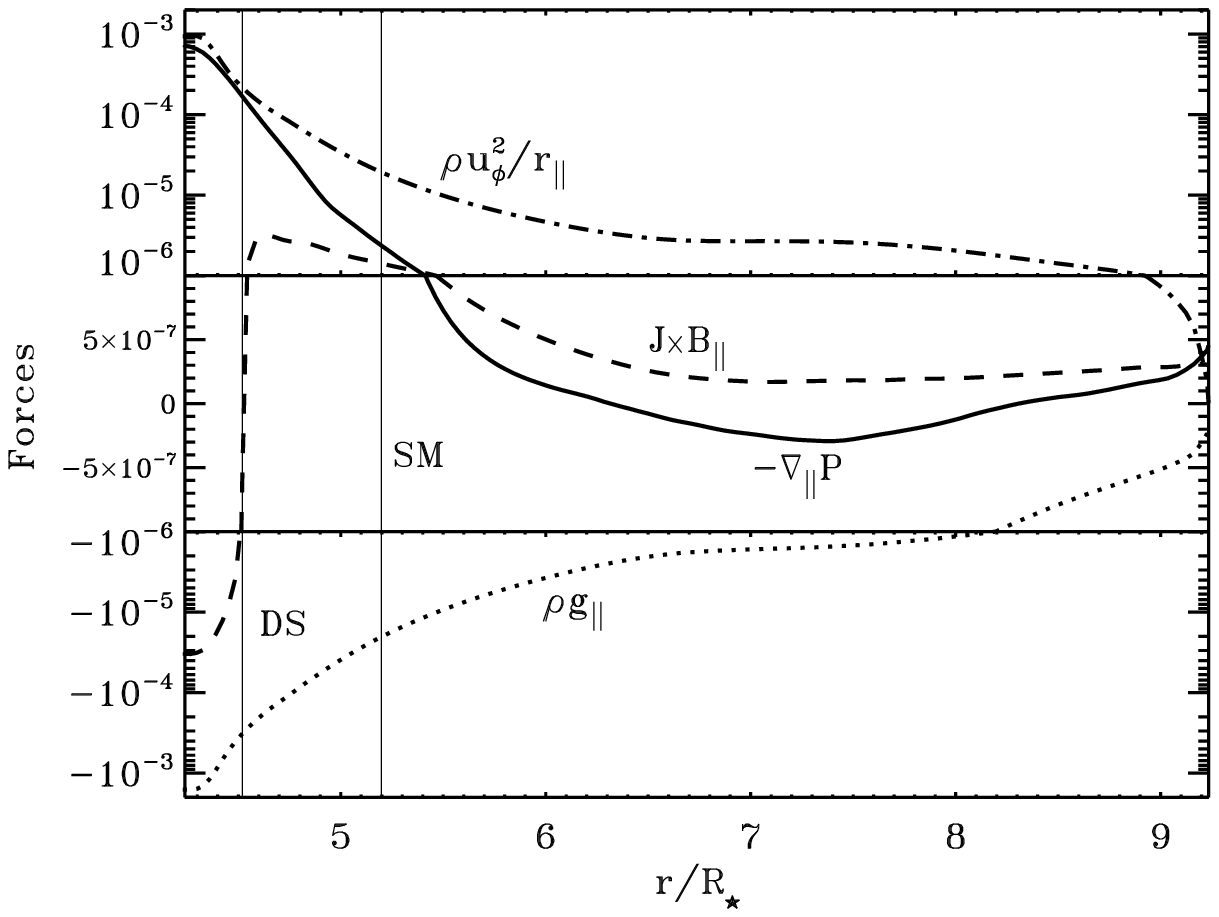}
   \includegraphics[width=0.5\textwidth]{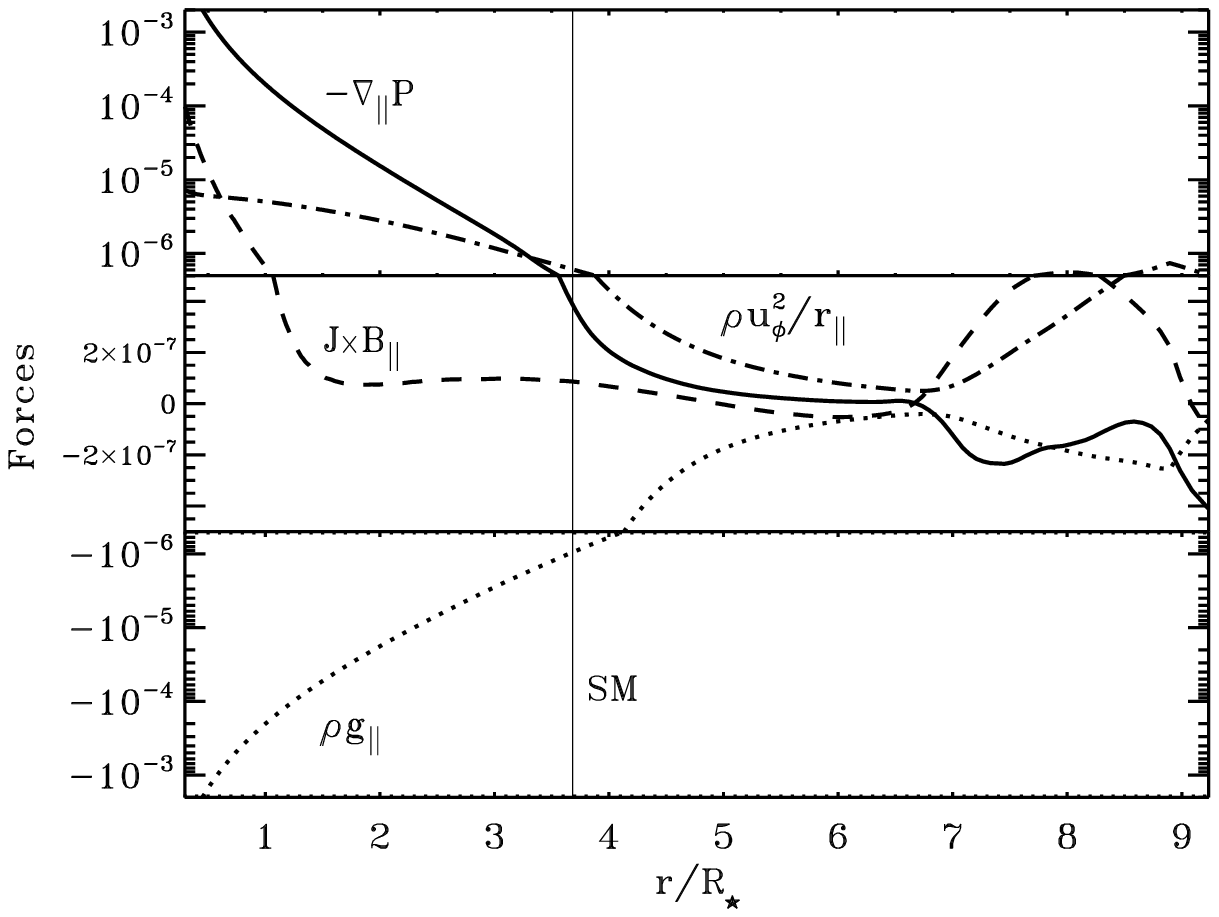}
  \caption{Forces acting along a field line connecting the star and the disk that is mass-loaded by the MEs in case C03. The left panel refers to the forces acting from the disk
                  midplane up to the cusp of the magnetic surface, while the right panel shows the forces from the stellar surface up to the cusp. We selected a magnetic
                  surface anchored at $4.3 R_\star$ at the disk midplane. The plots have been obtained by time averaging the forces over 54 stellar rotation periods. 
                  We plot the thermal pressure gradient ({\it solid line}), the centrifugal term ({\it dot-dashed line}), the Lorentz force ({\it dashed line}) and the
                  gravitational pull ({\it dotted line}). Vertical lines mark the position of the disk surface (DS) and the slow-magnetosonic point (SM).}
  \label{fig:forcesmej}
  \end{figure*}  
    
\subsubsection{Forces}
\label{sec:forces}

To complete the analysis of the dynamical properties of the MEs for the reference case C03, we take into account the forces that drive these outflows. 
Since mass is accelerated both from the disk and the star, we consider an inflating field line anchored in the disk at $R= 4.3 R_\star$ still connecting 
the disk and the star: we plot the component of the poloidal forces parallel to this field line from the disk midplane up to the field line cusp (left panel in Fig. \ref{fig:forcesmej})
and from the stellar surface up to the cusp (right panel in Fig. \ref{fig:forcesmej}) to analyze the driving mechanism of the MEs from the disk and from
the star respectively.  The forces have been obtained from a time-averaged snapshot, in order to smooth out transient features.

In the left panel of Fig. \ref{fig:forcesmej} we show that the acceleration of the mass of the MEs coming from the disk is largely due to a combination of 
centrifugal and magnetic effects, as in a typical disk-driven outflow. In addition, the thermal pressure gradient $-\nabla_\parallel P$ is comparable 
to the centrifugal acceleration at the disk surface (DS, defined as the point where the Lorentz force $\vec{J}\times\vec{B}_\parallel$ changes sign) 
and crucially contributes to the outflow acceleration. As it will be shown in Section \ref{sec:disktor}, this enhanced thermal pressure gradient is due to 
the push of the accretion flow against the magnetospheric wall and is responsible for the high mass load of the MEs coming from the disk.
Thermal effects are important since the centrifugal term $\rho u_\phi^2/r_\parallel$ is not sufficient to counteract the gravitational pull 
$\rho g_\parallel$ at the base of the flow: as it will be more extensively discussed in Section \ref{sec:disktor}, in the disk acceleration region of the 
MES, the disk rotation becomes strongly sub-Keplerian.

The right panel of Fig. \ref{fig:forcesmej} shows that the pressure gradient provides the initial thrust of the mass of the MEs coming from the star.
Even if this term is most probably not of thermal type, as it is assumed in our simulations, it is needed to drive any kind of stellar outflow from slowly
rotating stars, where magneto-centrifugal effects are not sufficient to give the initial push. The sudden change in the profile of the forces at $r \approx
7 R_\star$ happens when the material accelerated from the star comes across the mass coming from the disk. The latter is characterized by a 
higher density (see, for example, the change in the centrifugal push and gravitational pull), confirming the fact that the MEs inertia is dominated by
the mass loaded from the disk, as already discussed in Section \ref{sec:mflux}.
The profile of the Lorentz force has some interesting features. We first recall that the Lorentz force parallel to a field line in the poloidal plane
is related to the toroidal component of the force, according to the relation:
\[
\left( \vec{J} \times \vec{B} \right)\cdot \vec{B}_\mathrm{p} = -  \left( \vec{J} \times \vec{B} \right)\cdot \vec{B}_\phi \; ,
\]
which shows that a Lorentz force accelerating (braking) along a field line also accelerates (brakes) in the toroidal direction. 
Therefore we can see that the mass loaded from the star is subject to a toroidal acceleration close to the stellar surface while, getting closer to the
part of the MEs coming from the disk, between $5R_\star < r < 7R_\star$, it is spun-down. This clearly indicates that the star is trying to spin-up
the plasma attached to this field line, therefore loosing angular momentum, countering the material of the MEs coming from the disk 
that is trying to spin it down. 

\subsection{Dynamical properties of  stellar winds and disk winds}
\label{sec:swdw}

In our simulations, a stellar wind is accelerated along the open field lines anchored into the stellar surface. 
Due to the energetic limitations illustrated in the Introduction, we limit the mass outflow rate of the stellar wind to a few percent
of the mass accretion rate, so that the energy needed to drive initially the outflow corresponds to a small fraction (less than $10\%$) 
of the power dissipated by accretion onto the stellar surface. In Fig. \ref{fig:Alpha03rates} it is also plotted the temporal evolution of the
stellar wind mass outflow rate, calculated as:
\begin{equation}
\dot{M}_\mathrm{SW} = 4\pi R_\star^2 \int^{\theta_a}_{0} \rho u_R \sin\theta \, \mathrm{d}\theta \; ,
\label{eq:mwind}
\end{equation}
where $\theta_a$ corresponds to the anchoring angle of the last open stellar magnetic surface: the outflow rate corresponds
on average, to $\approx 1.6\%$ of the mass accretion rate measured onto the surface of the star (Eq. \ref{eq:maccs}). Comparing the mass
outflow rate of the reference case with the stellar wind of case E1, we find that the mass loss rate of case C03 corresponds on average 
to $\sim30 \%$ of the outflow rate of case E1, despite in the two cases we assumed approximately the same injection speed, density and
temperature. The mass ejection rate difference is therefore due to a different size of the launching area: 
Table \ref{table:fluxes} shows that in cases C03 and E1 the closed magnetosphere contains approximately the same amount of stellar
flux ($\Phi_\mathrm{MC}+\Phi_\mathrm{CD}$), even if in case C03 this flux is compressed in a much smaller region closer to the star
(see $R_\mathrm{cm}$ in Table \ref{table:fluxes}); the presence of the MEs, which are almost absent in case E1, reduces the amount of
magnetic flux and stellar surface which is available to launch the stellar wind. This examples shows that a 
self-consistent model of stellar winds from accreting protostars must take into account the presence of the accretion funnels, which can strongly 
affect the geometry of the launching region and the morphology of the magnetic surfaces along which the wind flows. 

We also estimated the specific angular momentum extracted by the stellar winds from the star (lower panel in Fig. \ref{fig:Alpha03lever}) 
defined as:
\begin{equation}
j_\mathrm{SW} = \frac{\dot{J}_\mathrm{SW}}{\dot{M}_\mathrm{SW}} = \overline{r}_\mathrm{A}^2 \Omega_\star \; ,
\label{eq:jsw}
\end{equation}
where the stellar wind torque $\dot{J}_\mathrm{SW}$ has been obtained by integrating Eq. (\ref{eq:jflux}) over the stellar surface
from which the wind is launched, as in Eq. (\ref{eq:mwind}). This last equation also provides the definition of the average magnetic
lever arm $\overline{r}_\mathrm{A}$. The average specific angular momentum of the case taken into account, $j_\mathrm{SW} 
\approx 44 \sqrt{GM_\star R_\star}$, corresponds to a lever arm $\overline{r}_\mathrm{A} \approx 21 R_\star$. 
Notice that this value is slightly larger than the value found in case E1 ($\overline{r}_\mathrm{A} \sim 19 R_\star$). Since the mass to magnetic 
flux ratio ($\eta = \rho v_\mathrm{p}/B_\mathrm{p}$) of the stellar winds  of these two cases are comparable, the different topology of the magnetic 
surfaces due to the interaction with the accretion funnels and the MEs likely determines the different lever arms. 
We just point out that the magnetic configuration found in case E1 allows wide opening winds, while in case C03 
the stellar wind assumes a more conical shape, where the presence of the MEs focuses the open magnetic flux towards the rotation axis. 

We can estimate the energy content of the stellar wind by evaluating Eq. (\ref{eq:invaren}) at the stellar surface. It can be shown that this
expression can be rewritten as:
\begin{equation}
\frac{e_\mathrm{SW}}{V_{\mathrm{K}\star}^2} = \left(\frac{\overline{r}_\mathrm{A}}{R_\star}\right)^2\delta_\star^2 - 
\frac{1}{2}\overline{\sin^2\theta}\,\delta_\star^2+\frac{h_\star}{V_{\mathrm{K}\star}^2}-1 \; .
\end{equation}
The first term in the right hand side, corresponding to the magnetic flux, is the dominant one. Besides, the stellar boundary conditions on the outflowing 
material impose that $h_\star \approx V_{\mathrm{K},\star}^2$, so that the enthalpy provides the initial drive to the stellar wind: even if this term must 
be of different origin, it is just meant to mimic the effect of an extra pressure term.
Therefore, we obtain that $e_\mathrm{SW} \approx 4.4 V_{\mathrm{K},\star}^2$. 
This specific energy would correspond to a maximum asymptotic speed $u_{\mathrm{p}, \infty} \approx 2.9 V_{\mathrm{K}\star}$, but the outflow
has attained a poloidal speed $\approx  V_{\mathrm{K}\star}$ at the end of the computational domain. The stellar wind has therefore the potential to be
a light and very fast outflowing component, provided an efficient magnetic-to-kinetic energy conversion can be attained and this crucially depends 
on the asymptotic magnetic flux distribution. Notice that the total specific energy of the stellar wind is much higher than the one of the MEs 
(see Fig. \ref{fig:Invarmej}), almost one order of magnitude, mostly since the MEs are much more massive and therefore have less energy per particle 
available. This translates in a lower limit on the maximum terminal speed achievable by the MEs.

We are not going to describe in great detail the properties of the disk wind accelerated along the open magnetic surface threading the accretion disk. 
Its mass outflow rate, obtained by integrating the mass flux equation Eq. (\ref{eq:mflux}) at the disk surface beyond $R_\mathrm{out}$, is rather small,
see Fig. \ref{fig:Alpha03rates}, and mostly concentrated close to radius $R_\mathrm{out}$, where the magnetic field is stronger. 
Besides, it is clear from Fig. \ref{fig:globalview} that the Alfv\'en surface lies very close to the disk: this means that the disk wind is characterized 
by a rather small  Alfv\'en radius, extracts a limited amount of angular momentum from the disk and therefore has a negligible impact on the angular 
momentum distribution of the star-disk system. This will be shown more quantitatively in the following Section.

\subsection{The disk angular momentum}
\label{sec:disktor}

 \begin{figure}
 \centering
 \resizebox{0.9\hsize}{!}{\includegraphics*{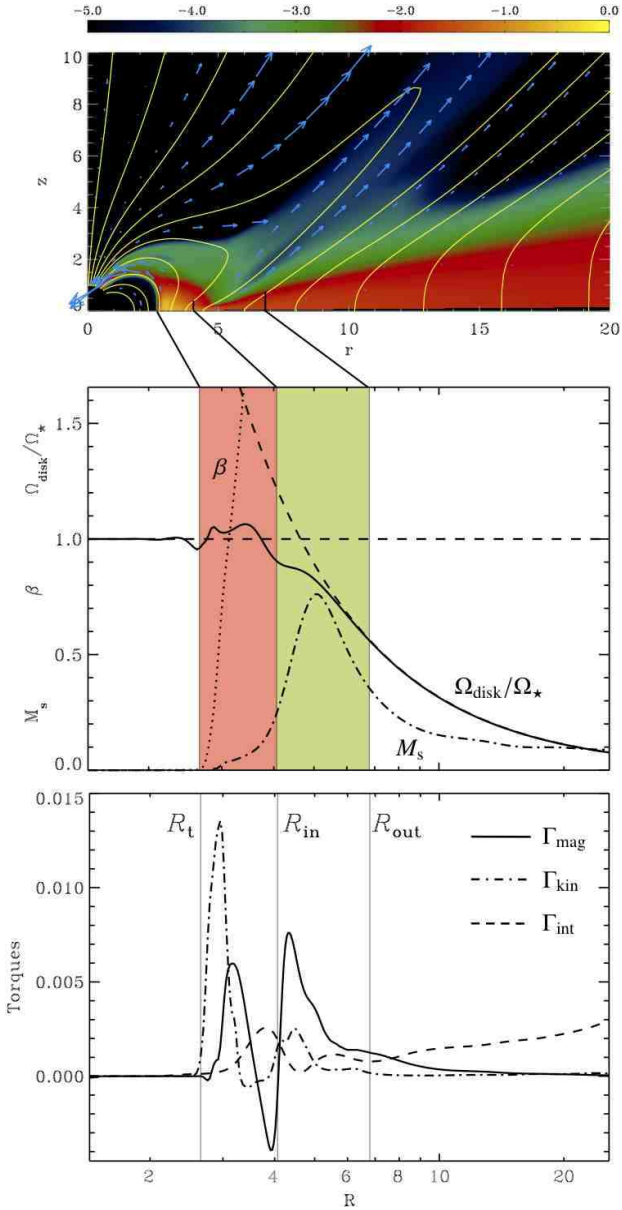}}
 \caption{{\it Upper panel}. Logarithmic density map with sample field lines (yellow solid lines) and speed vectors (blue arrows)
                  superimposed. {\it Middle panel}. Radial profiles at the disk midplane of the rotation speed of the accretion disk 
                  $\Omega/\Omega_\star$ ({\it solid line}), accretion sonic Mach number ({\it dot-dashed line}) and plasma $\beta$ 
                  ({\it dotted line}). The Keplerian and the $\Omega = \Omega_\star$ rotation profiles are plotted with a dashed line.
                  {\it Lower panel}. Radial profiles of the specific torques acting on the disk (see the text for definitions): 
                  magnetic ($\Gamma_\mathrm{mag}$, {\it solid line}), internal ($\Gamma_\mathrm{int}$, {\it dashed line} and kinetic 
                  ($\Gamma_\mathrm{kin}$, {\it dot-dashed line}) torques. In the three panels we marked with vertical lines 
                  the position of the truncation radius ($R_\mathrm{t}$) and the anchoring radii at the disk surface of the outermost
                  magnetic surface steadily connecting the star and the disk ($R_\mathrm{in}$) and of the innermost open field line
                  threading the disk ($R_\mathrm{out}$).  The panels represent temporal averages over 54 stellar rotation periods.}       
 \label{fig:Comp03}
 \end{figure}
 
In this Section we analyze the impact of disk-driven outflows, MEs and disk winds, on the angular momentum distribution of the 
accretion disk. We start our analysis by taking into account the torques acting on the circumstellar disk and their effects on the 
disk dynamics.  Accretion can be triggered (or hampered) by internal torques, e.g. turbulent viscosity, redistributing 
angular momentum radially inside the disk, or external torques, magnetic stresses and mass transfers, extracting 
or supplying angular momentum at the disk surface. This balance can be expressed in a steady situation by 
considering the angular momentum conservation inside an annulus of the disk of radial width $\mathrm{d}r$ and 
thickness $2H$. Following the notation of \citetalias{ZanFer09}, the angular momentum conservation can be formulated as:
\begin{equation}
\Gamma_\mathrm{acc} = \Gamma_\mathrm{int} + \Gamma_\mathrm{mag} +\Gamma_\mathrm{kin} \; , 
\label{eq:amtrans}
\end{equation}
where 
\begin{equation}
\Gamma_\mathrm{acc} = \frac{\mathrm{d}}{\mathrm{d} r} \left( \dot{M}_\mathrm{acc} r^2 \Omega_\mathrm{disk} \right) 
\end{equation}
gives the angular momentum advection through the annulus. The mass accretion rate $\dot{M}_\mathrm{a}$ 
is defined as:
\begin{equation}
\dot{M}_\mathrm{acc} =  - 2 \pi r \int_{-H}^{+H} \rho u_r  \, \mathrm{d}z \; .
\label{eq:maccd}
\end{equation}
We express the internal torque $\Gamma_\mathrm{int}=\Gamma_\mathrm{visc}+\Gamma_\mathrm{B}$ 
as the sum of the viscous ``turbulent'' torque:
\begin{equation}
\Gamma_\mathrm{visc} =  -2 \pi \frac{\mathrm{d}}{\mathrm{d} r} \left( r^2 \int_{-H}^{+H} \tau_{r \phi} \, \mathrm{d}z \right) 
\end{equation}
and the radial magnetic transport:
\begin{equation}
\Gamma_\mathrm{B} =  -\frac{\mathrm{d}}{\mathrm{d} r} \left( \frac{r^2}{2} \int_{-H}^{+H} B_\phi B_r \, \mathrm{d}z \right) \; .
\end{equation}
We included this last term for the sake of completeness: nevertheless, the internal torque $\Gamma_\mathrm{int}$ is 
dominated by the viscous term. 
The torque exerted by the large-scale magnetic field, extracting angular momentum at the disk surface, is defined as:
\begin{equation}
\Gamma_\mathrm{mag} =  r^2 B_\phi^+ B_\mathrm{p}^+ \; ,
\label{eq:gmag}
\end{equation}
The kinetic torque, determined by the mass exchange at the disk surface, is given by:
\begin{equation}
\Gamma_\mathrm{kin} =   -4 \pi r^3 \rho^+ \Omega^+u_p^+ \; ,
\label{eq:gkin}
\end{equation}
According to these definitions, a positive torque in the right hand side of Eq.~(\ref{eq:amtrans}) extracts angular 
momentum from the annulus, thus favoring accretion. 

In Fig. \ref{fig:Comp03} we show the the right hand 
side torques of Eq. (\ref{eq:amtrans}) as a function of the radius $r$ (lower panel), in the middle panel the effect of these torques on 
the disk structure (disk angular speed, accretion sonic Mach number, $\left.M_\mathrm{s} = \left|u_r\right|/\sqrt{P/\rho}\right|_{z=0}$ 
and plasma beta, $\left.\beta = 8\pi B^2/P\right|_{z=0}$), while in the upper panel we display the corresponding density maps with 
field and stream lines superposed. The three panels are not snapshots at 
a given time, but have been obtained by averaging over 54 rotation periods of the star, from time $t =38$ up to $t=92$, 
in order to smooth out possible transient features. Besides, we marked three radii 
corresponding to the truncation  radius ($R_\mathrm{t}$), the anchoring radius onto the disk surface 
of the outermost magnetic field line steadily connecting the disk with the star ($R_\mathrm{in}$) and 
of the innermost open field line threading the disk ($R_\mathrm{out}$). 
These radii allow to distinguish three different zones: in the region between $R_\mathrm{t}$ and 
$R_\mathrm{in}$, shaded in red in the middle panel, the disk can directly exchange angular momentum with the 
star and form the accretion columns; the region between $R_\mathrm{in}$ and $R_\mathrm{out}$, shaded in
green in the middle panels, despite being magnetically linked to the star, does not exchange angular momentum directly with it, 
since the magnetic surfaces have expanded too much to be causally connected directly with the star. The material ejected 
from this region escapes the stellar potential well instead of being accreted and the angular 
momentum extracted at the disk surface is transferred to the outflowing material: as already pointed out, this is the region 
from which the mass coming from the disk is accelerated to fuel the MEs. The disk region outside $R_\mathrm{out}$ is threaded 
by open field lines along which, depending on the magnetic field strength, a disk wind can be accelerated. 
 
The lower panel shows that the internal turbulent transport is responsible for driving accretion on a large scale ($r \gtrsim 8 \, R_\star$). 
Even if the disk is threaded by the magnetic flux left by the opening of the dipolar magnetosphere, at this distance from the star the 
large-scale field is too weak to accelerate a powerful enough disk wind to exert a noticeable torque. Getting closer to the star, 
the large-scale open magnetic field threading the disk starts to be strong enough to 
influence the accretion dynamics: starting already in the region outside $R_\mathrm{out}$, the magnetically-driven disk wind  
increasingly contributes to drive the accretion flow. Correspondingly, the disk dynamics start to change behavior (middle panels): 
the rotation profile is still Keplerian but, due to the growing magnetic torque, the accretion Mach number increases towards 
trans-sonic values. Since these outflows do not affect the disk angular momentum distribution in a relevant way 
(the rotation stays Keplerian) and extract a negligible amount of accreted mass, they will not be discussed in greater detail.

The magnetic torque progressively becomes dominant in the disk region connected to the star from which the 
MEs arise ($R_\mathrm{in} < R < R_\mathrm{out}$). The middle panel shows that, in this region, the 
accretion Mach number starts to decrease after it has attained a maximum, almost sonic value: this corresponds to an adiabatic 
compression due to the push of the accretion flow against the stellar magnetosphere which is acting as a magnetic wall. This 
compression determines the enhanced mass-loading of the magnetospheric ejections, as the kinetic torque curves show clearly
in the lower panel.  As already noticed in \citetalias{ZanFer09}, this same effect is crucial to mass-load the accretion funnels. 
Besides, the ejection torque (kinetic plus magnetic) becomes strong enough so that in this region the disk rotation becomes 
sub-Keplerian and even sub-stellar.  In order to provide a more precise idea of the  amount of angular momentum extracted by 
the MEs from the star-disk system, we can consider the angular momentum flux carried by the accretion flow through a vertical 
section of the disk at $R_\mathrm{out}$:
\begin{equation}
\dot{J}_{\mathrm{acc},R_\mathrm{out}} = \left. -2\pi r^2 \int_{-H}^{+H} \left( \rho u_r u_\phi - \tau_{r\phi} -\frac{B_\phi B_r}{4\pi} \right)  \, \mathrm{d}z  \, \right|_{R_\mathrm{out}} \; .
\label{eq:jaccdisk}
\end{equation}
Without any other interaction of the disk with the surroundings inside $R_\mathrm{out}$, this would be the spin-up torque exerted 
by the accretion flow onto the star. We plot in the upper panel of Fig. \ref{fig:Alpha03lever} the time evolution of the specific angular momentum carried through 
$R_\mathrm{out}$:
\begin{equation}
j_{\mathrm{acc},R_\mathrm{out}}= \frac{\dot{J}_{\mathrm{acc},R_\mathrm{out}}}{\dot{M}_{\mathrm{acc},R_\mathrm{out}}} \; ,
\label{eq:jaccdisksp}
\end{equation}
where the disk accretion rate $\dot{M}_{\mathrm{acc},R_\mathrm{out}}$ has been obtained by evaluating Eq. (\ref{eq:maccd}) at $R_\mathrm{out}$. 
The specific angular momentum $j_{\mathrm{acc},R_\mathrm{out}}$ is  equal, on average, to $j_{\mathrm{acc},R_\mathrm{out}} = 1.8 \sqrt{GM_\star R_\star}$.  
Notice that, since the definition of $\dot{J}_{\mathrm{acc},R_\mathrm{out}}$ (Eq. \ref{eq:jaccdisk}) includes the viscous and magnetic torques, 
$j_{\mathrm{acc},R_\mathrm{out}}$ (Eq. \ref{eq:jaccdisksp}) is lower than the specific angular momentum expected from the advection process only 
($\approx \sqrt{GM_\star R_\mathrm{out}} \approx 2.6 \sqrt{GM_\star R_\star}$).
Using the estimates done in Section \ref{sec:mes}, we can see that the MEs extract directly from the disk a fraction 
$\left(\dot{M}_\mathrm{ME,d}j_\mathrm{ME,d}\right)/ \left(\dot{M}_{\mathrm{acc},R_\mathrm{out}} j_{\mathrm{acc},R_\mathrm{out}}  \right) \approx 53\%$ 
of the angular momentum carried by the disk through $R_\mathrm{out}$, forcing the accretion disk to rotate at a sub-stellar rate. 
These ejections have therefore the important effect of extracting a relevant fraction of the disk angular momentum and consequently reducing the accretion torque. 
Besides, the material ejected from the disk to fuel the MEs clearly rotates slower than the star: since the mass ejected from the disk is also magnetically connected
to the star and largely dominates the MEs inertia, it can extract angular momentum from the star thanks to a differential rotation effect. 

The disk region inside $R_\mathrm{in}$ can exchange its angular momentum directly with the star. Therefore, all the angular 
momentum that flows in across $R_\mathrm{in}$  is eventually accreted by the star and thus determines a spin-up torque.
In this magnetically connected region, the stellar rotation tries to force the disk to co-rotate with it. Since in this example 
the disk rotates slower than the star at $R_\mathrm{in}$, the stellar rotation spins-up the disk back to $\Omega_\star$,
thus exerting a negative magnetic torque on the disk (see the $\Gamma_\mathrm{mag}$ curve in the lower panel of Fig. 
\ref{fig:Comp03}) and extracting a fraction of the stellar angular momentum back to the disk. 
Notice that the $\Gamma_\mathrm{mag}$ curve becomes positive again close to the truncation region and the 
disk starts to transfer angular momentum to the star.  The truncation region is dominated by the kinetic torque, due to the mass 
loaded onto the base of the accretion funnels\footnote{For a discussion on the dynamics of the truncation region and the accretion 
funnels we refer the reader to \citet{Bessolaz08} and \citetalias{ZanFer09} (Section 3.1).}. The external torque (magnetic plus kinetic) 
is characterized by a double-peaked profile, the inner positive peak being associated with the star-disk angular 
momentum exchange and the outer one with the torque exerted on the disk by the magnetospheric ejections. 
       
\subsection{The stellar angular momentum}
\label{sec:sangmom}

 \begin{figure}[!t]
 \resizebox{\hsize}{!}{\includegraphics*{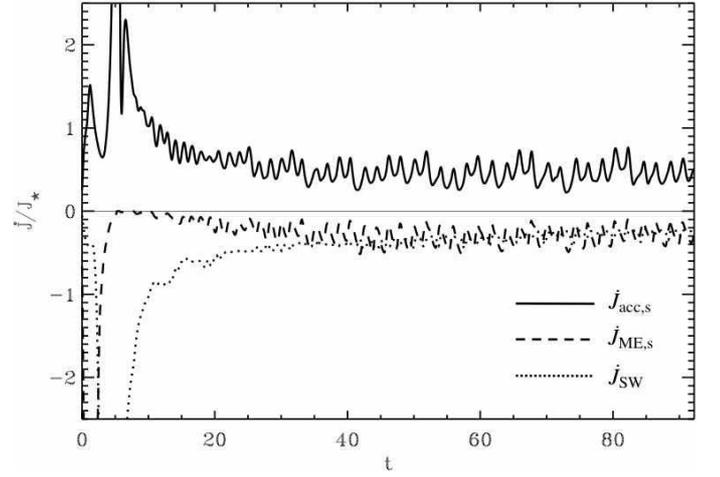}}
 \caption{Temporal evolution of the torques acting directly onto the star, normalized to the stellar angular momentum. Plotted are the
                 accretion torque ({\it solid line}), the stellar wind torque ({\it dotted line}) and the torque exerted by the MEs onto the stellar surface
                 ({\it dashed line}). Conventionally a positive (negative) torque spins up (down) the stellar rotation.}
 \label{fig:Alpha03torques}
 \end{figure}
 
In this Section we evaluate the impact of accretion and stellar outflows (MEs and stellar winds) on the temporal evolution of the 
angular momentum of the central star. We can express the time derivative of the stellar angular momentum $J_\star$ by integrating the 
angular momentum conservation equation over the stellar volume:
\begin{equation}
\frac{\mathrm{d} J_\star}{\mathrm{d} t} = \dot{J}_\mathrm{acc,s} + \dot{J}_\mathrm{ME,s} +  \dot{J}_\mathrm{SW} \; ,
\label{eq:jstar}
\end{equation}
where we separated the contributions to the torque due to accretion, magnetospheric ejections and stellar winds. 
A positive (negative) angular momentum flux in the right hand side of Eq. (\ref{eq:jstar}) correponds to a spin-up (spin-down) torque.
The three torques have been obtained by integrating Eq. (\ref{eq:jflux}) over three different parts of the stellar surface:  the area threaded 
by open field lines to evaluate the stellar wind torque $\dot{J}_\mathrm{SW}$ (see Section \ref{sec:swdw}); 
the area threaded by magnetic surfaces that undergo periodic inflation/reconnection events to evaluate the MEs contribution 
$\dot{J}_\mathrm{ME,s}$ (see Section \ref{sec:mes}); the area threaded by field lines
steadily connecting the star with the disk to evaluated the accretion torque $\dot{J}_\mathrm{acc,s}$. 
As already pointed out in \citetalias{ZanFer09}, the contribution of the kinetic terms is completely negligible at the stellar surface, both for accreting and 
outflowing components, which are completely dominated by the magnetic torque. In the following discussion we will consider the total torques,
keeping in mind that the magnetic contribution is prevailing. In Fig. \ref{fig:Alpha03torques} we show the temporal evolution of the torques acting on the star, 
normalized to the stellar angular momentum $J_{\star} = I_{\star} \Omega_\star$, where $I_{\star} = k^2 M_{\star} R_{\star}^2$ is the stellar moment 
of inertia. We assumed the typical normalized gyration radius of a fully convective star, i.e. $k^2=0.2$. Using this normalization, the curves in 
Fig. \ref{fig:Alpha03torques} provide directly the inverse of the characteristic braking (or speed-up) timescale. To retrieve the physical units, the plots must be
multiplied by Eq. (\ref{eq:tnormdens}) that, expressed in terms of the stellar magnetic field intensity, takes the form:
\begin{equation}
\left.\frac{\dot{J}}{J_\star}\right|_0 = 10^{-6} \,  \left( \frac{B_\star}{1\, \mathrm{kG}}\right)^{2} \left( \frac{M_\star}{0.5 \, M_\odot}\right)^{-3/2} \left( \frac{R_\star}{2 \, R_\odot}\right)^{5/2} \; \mathrm{yr}^{-1} \; .
\label{eq:tnorm}
\end{equation}  

\subsubsection{Accretion torque}
\label{sec:actor}

A rotating accretion disk is likely to exchange angular momentum besides mass onto the central object, thus providing a spin-up torque to the latter.
The accretion spin-up torque can be parametrized as $\dot{J}_\mathrm{acc,s} = \dot{M}_\mathrm{acc,s} j_\mathrm{acc,s}$, where 
$j_\mathrm{acc,s}$  is the specific angular momentum transported by the accretion streams and also along  the magnetic surfaces connected to the disk
which are not mass-loaded. 
A common parametrization for the specific accreted angular momentum  is $j_\mathrm{acc} = \sqrt{GM_\star R_\mathrm{t}}$, implying that a Keplerian 
accretion disk transfers to the star the angular momentum possessed in the truncation region. In our simulations we can estimate the 
accuracy of this approximation. For example the specific angular  momentum transferred by the disk to the star in case E1 from \citetalias{ZanFer09}
is larger than this reference value,  $j_\mathrm{acc,s} \sim 1.2 \sqrt{GM_\star R_\mathrm{t}}$: since in this case the stellar magnetosphere 
is connected to the accretion disk over a large extent, even beyond the corotation radius, the star can extract angular momentum from the disk in the
entire region from $R_\mathrm{t}$ up to $R_\mathrm{co}$.
On the other hand, we already pointed out that in the ``compact'' magnetic configuration depicted in Fig. \ref{fig:globalview}, 
the accretion torque is given by the angular momentum flux through a disk surface from $R_\mathrm{t}$ up to $R_\mathrm{in} < R_\mathrm{co}$.
In case C03, the average accreted angular momentum is approximately equal to $60\%$ of the reference Keplerian value  
$\sqrt{GM_\star R_\mathrm{t}}$.  As already pointed out in Section \ref{sec:disktor}, this effect is due the presence of MEs, which are extracting 
a consistent fraction of the angular momentum of the accretion disk before it is accreted onto the star. 
   
\subsubsection{Stellar wind torque}
\label{sec:swind}

In Section \ref{sec:swdw} we estimated the mass outflow rate and the specific angular momentum extracted by the stellar wind from the star in the reference
case C03. We can now compare the stellar wind torque with the accretion torque. On average, the stellar wind spin-down torque extracts around $66\%$ of the
accretion torque. This efficiency seems to be rather high, given the low ejection efficiency of the wind. 
For example, we recall that the stellar wind of case E1 from \citetalias{ZanFer09}, was able to balance $20\%$ of the accretion torque only. Three effects have enhanced 
the efficiency of the spin-down torque: the mass ejection efficiency in case C03 is slightly higher than in case E1 ($1.6\%$ vs. $1.2\%$); in case C03 the 
magnetic lever arm is larger ($\overline{r}_\mathrm{A} \approx 21 R_\star$ vs. $\overline{r}_\mathrm{A} \approx 19 R_\star$). But the most important and 
interesting effect is due to the fact that in case C03 the MEs have already extracted a substantial amount of the disk angular momentum, reducing the 
accretion torque by a factor around $50\%$ (see Section \ref{sec:disktor}) and therefore enhancing the efficiency of the stellar wind torque.

\subsubsection{Magnetospheric ejections torque}
\label{sec:majstar}

As we showed in Section \ref{sec:mes}, besides reducing the accretion spin-up torque, MEs are able to exchange angular momentum directly with the star.
As we have already shown, the angular momentum exchange with the star is mainly controlled by the differential rotation between the star and the MEs:
if at the cusp of the field line the mass loaded from the disk rotates slower (faster) than the star, the MEs exert a spin-down (spin-up) torque.
Consistently, since we find that in case C03 MEs rotate slower than the star (see Section \ref{sec:disktor}), they exert a net spin-down torque, 
equal to $\approx 74\%$ of the accretion torque. 

Summarizing, in the case C03 the combined action of MEs and stellar winds is able to balance the spin-up due to the accretion torque. Notice that,
even if MEs represent the dominant effect, extracting globally around $88\%$ of the disk angular momentum accreted through $R_\mathrm{out}$,
the stellar wind can contribute significantly, extracting, in our example, around $31\%$ of the disk angular momentum flux through  $R_\mathrm{out}$.
Therefore, in the specific example analyzed in the previous Sections, the star is subject to a net spin-down torque: using Eq. (\ref{eq:tnorm}) to
normalize to physical units the sum of the three torques plotted in Fig. \ref{fig:Alpha03torques}, we can estimate a characteristic spin-down timescale
around $5.5\times10^6$ yr.

\section{Varying the mass accretion rate}
\label{sec:othercases}
In this Section we analyze cases C1 and C01 and compare them with the results obtained from case C03, extensively presented in Section \ref{sec:mej}.
We recall that these two cases are characterized by a different disk viscosity coefficient $\alpha_\mathrm{v}$ (see Table \ref{table:cases}), with all the
other parameters of the problem left unchanged. 
Primarily, we use the $\alpha_\mathrm{v}$ parameter to change the mass accretion rate of the disk. 
On the other hand, this modifies also the magnetic Prandtl number $\mathcal{P}_\mathrm{m} = \nu_\mathrm{v}/\nu_\mathrm{m}$ of the disk. 
Therefore, bear in mind that our cases are not just characterized by different accretion rates and that other important aspects of the disk physics 
change. For example, the viscous accretion timescale varies from case to case, while
the different Prandtl number can have an impact on the inclination of the field lines at the disk surface and on the advection of the magnetic
flux in the part of the disk dominated by the viscous torque. Different accretion rates could also have been obtained by varying the disk 
density with a fixed Prandtl number, possibly giving somewhat different results.
 
\subsection{High accretion rate and stellar spin-up}
\label{sec:highrate}

 \begin{figure}
 \resizebox{\hsize}{!}{\includegraphics*{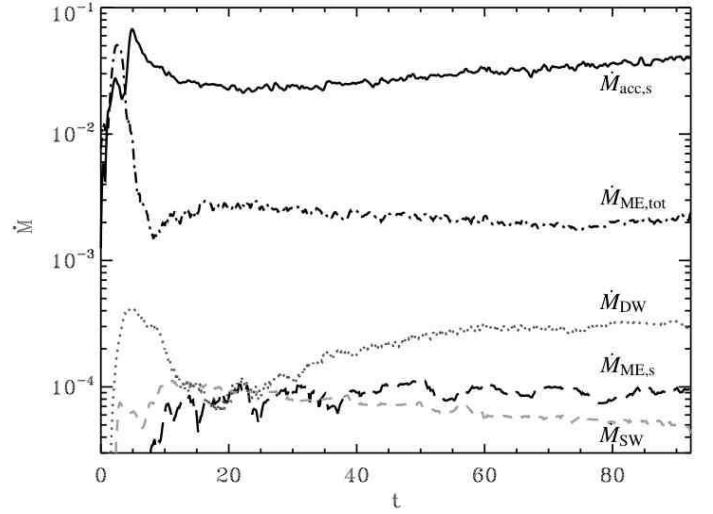}}
 \caption{Same as Fig. \ref{fig:Alpha03rates}, bur for case C1}
 \label{fig:Alpha1rates}
 \end{figure}
 
Case C1 is characterized by a value $\alpha_\mathrm{v} = 1$, determining an accretion rate around three times higher than the reference case C03,
as confirmed by the temporal evolution of the accretion rate measured onto the stellar surface plotted in Fig. \ref{fig:Alpha1rates}. There is another feature
that is worth noticing: while in case C03 the accretion rate attains a rather constant asymptotic value, in case C1 it slowly grows in time. We have already
shown in Section \ref{sec:disktor} that the accretion flow is controlled by the internal viscous torque far from the star, while it is determined by the external 
magnetic torques in the region of magnetospheric interaction. A mismatch between the two torques can trigger a long-term variability. For example, it seems
that in case C1 the magnetosphere cannot sustain the accretion rate of the outer disk. This determines a density and pressure buildup inside the disk that
causes, due to the $\alpha$ parametrization, an increase of the viscous torque and therefore of the accretion rate. The constant asymptotic value of the
accretion rate in case C03 is likely to be due to a better matching between the inner magnetic and outer viscous torques. 
 
 \begin{figure}
 \centering
 \resizebox{0.9\hsize}{!}{\includegraphics*{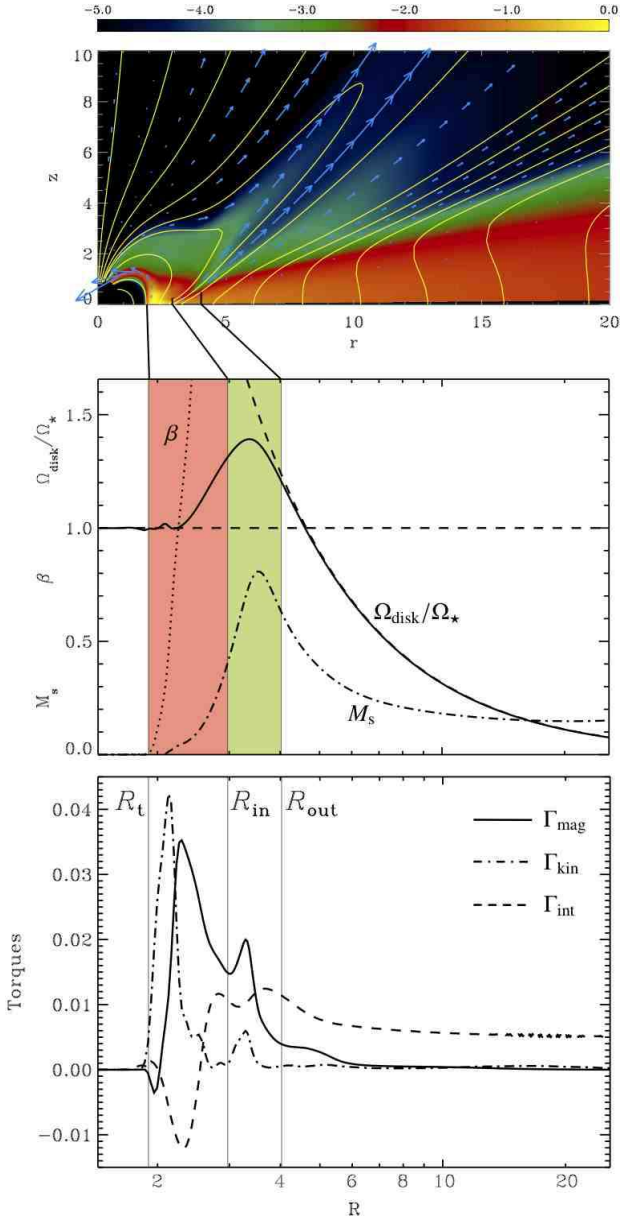}}
 \caption{Same as Fig. \ref{fig:Comp03}, but for case C1}
 \label{fig:Comp1}
 \end{figure}
  
\begin{figure}
 \resizebox{\hsize}{!}{\includegraphics*{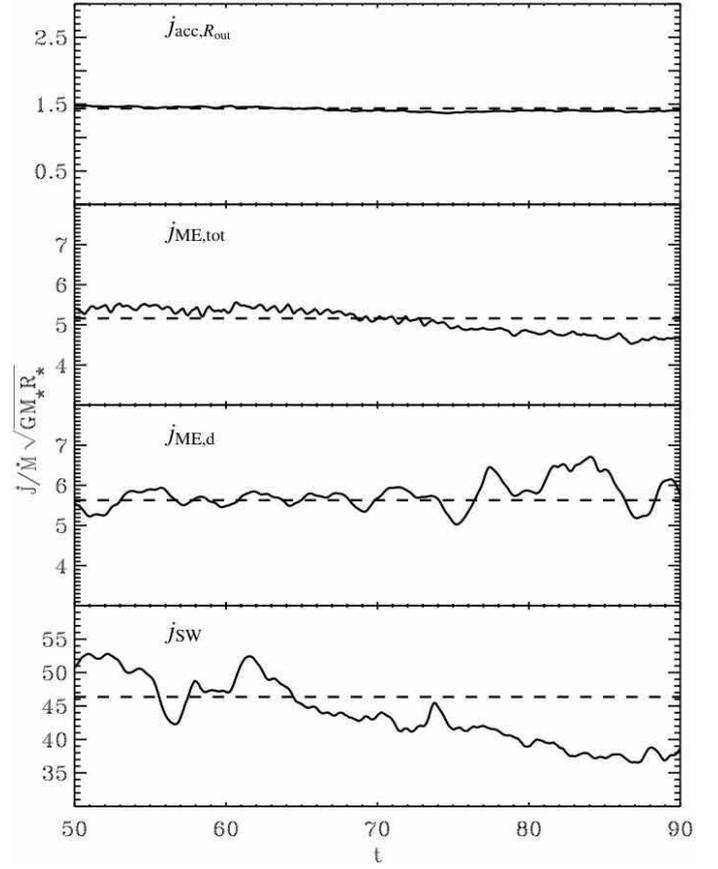}}
 \caption{Same as Fig \ref{fig:Alpha03lever}, but for case C1}
 \label{fig:Alpha1lever}
 \end{figure}

The global magnetic configuration of case C1 is very similar to case C03 (see Fig. \ref{fig:globalview}) and the same dynamical features can be found.
Regarding the MEs, their total mass outflow rate corresponds to around $6\%$ of the disk accretion rate and, as in case C03, the mass extracted
from the disk largely dominates the mass content of the MEs (see Fig. \ref{fig:Alpha1rates}). In the bottom panel of Fig. \ref{fig:Comp1} we can see that
the MEs are accelerated from a region of the disk located inside the Keplerian corotation radius which is more compact than in case C03, probably because 
of the larger magnetic Prandtl number. Outside $R_\mathrm{out}$, accretion is mainly triggered by the viscous torque, with a relatively small contribution
from a weak disk wind, which, nevertheless, accelerates the accretion flow towards an almost sonic accretion speed (middle panel in Fig. \ref{fig:Comp1}),
without modifying the Keplerian structure of the disk. Inside the launching region of the MEs ($R_\mathrm{in} < R < R_\mathrm{out}$), the torque due to the
MEs becomes dominant. In order to estimate the efficiency of this torque we calculated the specific angular momentum carried by the accretion flow
through $R_\mathrm{out}$ (see Eq. (\ref{eq:jaccdisk})) and the specific angular momentum extracted by the MEs from the disk (see Eq. (\ref{eq:jmed})) and
plotted them in the upper and third panel in Fig. \ref{fig:Alpha1lever} respectively:
MEs extract about $\left(\dot{M}_\mathrm{ME,d}j_\mathrm{ME,d}\right)/ \left(\dot{M}_{\mathrm{acc},R_\mathrm{out}} 
j_{\mathrm{acc},R_\mathrm{out}}  \right) \approx 20\%$ of the angular momentum carried by the disk through $R_\mathrm{out}$. The lower torque
efficiency measured in case C1 is due to a lower mass ejection efficiency but also to a smaller specific angular momentum extracted from the disk:
this is consistent with the fact that in case C1 the magnetic surfaces along which the MEs are accelerated have a larger inclination angle 
with respect to the disk surface (compare the upper panels of Fig. \ref{fig:Comp03} and \ref{fig:Comp1}), thus determining a smaller magnetic lever arm. 
Nevertheless, the MEs torque is strong enough so that the disk rotation speed becomes sub-Keplerian (middle panel in Fig. \ref{fig:Comp1}), 
but not sub-stellar as in case C03.
 
The angular momentum carried by the disk through $R_\mathrm{in}$ is finally transferred to the star and exerts a spin-up torque.
In the lower panel of Fig. \ref{fig:Comp1} it is possible to see that in the disk region directly connected to the star below $R_\mathrm{in}$ the angular
momentum is extracted from the disk and transferred to the star by a combination of magnetic and kinetic torques, where the peak of the latter corresponds 
to the angular momentum loaded onto the base of the accretion funnels. The external torque (magnetic plus kinetic) in the region $R< R_\mathrm{out}$
presents qualitatively the same double-peaked profile of case C03, with the inner peak corresponding to the angular momentum exchange with the star, while
the outer one representing the torque exerted by the MEs onto the disk. Using the same parametrization employed in Section \ref{sec:actor}, we can estimate that the
specific angular momentum transferred to the star corresponds to $j_\mathrm{acc,s} \approx 0.8 \sqrt{GM_\star R_\mathrm{t}}$ of the reference ``Keplerian'' value:
this clearly shows that also in case C1 one important effect of the MEs is to reduce the amount of angular momentum that the disk transfers to the star, although 
less efficiently than in case C03. 
    
\begin{figure}
 \resizebox{\hsize}{!}{\includegraphics*{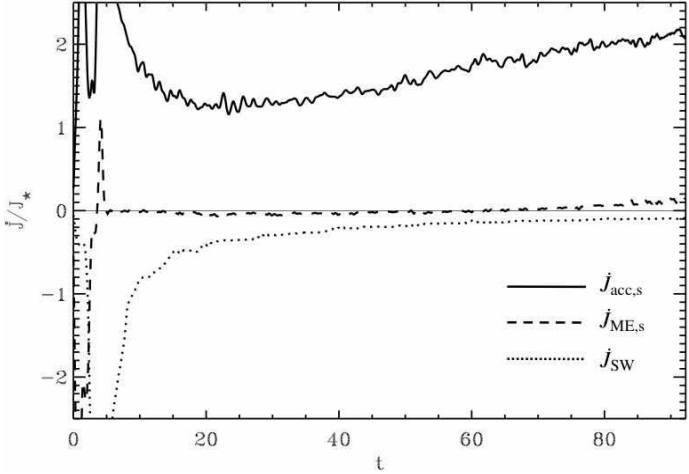}}
 \caption{Same as Fig. \ref{fig:Alpha03torques}, but for case C1}
 \label{fig:Alpha1torques}
 \end{figure}

The temporal evolution of the accretion torque is plotted in Fig. \ref{fig:Alpha1torques}, together with the torques exerted by the MEs and the stellar wind 
onto the stellar surface. The asymptotic increase of the accretion torque is clearly linked to the temporal growth of the mass accretion rate discussed previously.
The spin-down torque exerted by the stellar wind corresponds to $7.5\%$ of the accretion torque only. Despite carrying a specific angular momentum comparable
and even slightly larger than case C03 (see bottom panels of Fig. \ref{fig:Alpha03lever} and \ref{fig:Alpha1lever}), the low ejection rate 
($\dot{M}_\mathrm{SW} \approx 2\times10^{-3} \dot{M}_\mathrm{acc,s}$) strongly limits the efficiency of the spin-down torque. 
By inspecting Table \ref{table:fluxes}, it is possible to notice that in case C1 the connected magnetosphere contains more magnetic flux ($\Phi_\mathrm{CD}+\Phi_\mathrm{MC}$)
and it is even more compressed (see $R_\mathrm{cm}$) than in case C03: the accretion funnels 
are located at slightly higher latitudes, leaving less room for the acceleration area of MEs and stellar winds.

Besides, the stellar wind torque
slowly drops in time due to the decrease of the stellar wind specific angular momentum (lower panel in Fig. \ref{fig:Alpha1lever}): this is likely associated with the
increase of the mass accretion rate which can modify the shape of the magnetic nozzle at the base of the stellar wind, the acceleration efficiency of the outflow 
and  therefore its magnetic lever arm.
The torque exerted by the MEs directly onto the star is almost completely negligible. Besides, it is possible to see in Fig. \ref{fig:Alpha1torques} that for 
$t > 60$ it even becomes positive, contributing to the spin-up torque, even if to a very small extent. As we pointed out in Section \ref{sec:majstar}, the angular 
momentum exchange between the MEs and the star depends on a differential rotation effect between the star and the mass loaded from the disk onto the MEs.
We saw in Fig. \ref{fig:Comp1} that in case C1 the base of the MEs rotate at a sub-Keplerian but still super-stellar angular speed: consistently,  
the MEs do not directly brake the stellar rotation and can actually spin it up. The same effect is also shown in the two central panels of Fig. \ref{fig:Alpha1lever}:
the total specific angular momentum carried by the MEs (second panel) is smaller than the specific angular momentum extracted from the disk (third panel).
In case C1, contrary to case C03, the MEs can extract energy and angular momentum from the disk but then transfer a small fraction back to the star.
  
Since in this case the spin-down torques are very inefficient, the star experiences a strong spin-up torque due to accretion. Normalizing the torques plotted in 
Fig. \ref{fig:Alpha1torques} with Eq. (\ref{eq:tnorm}), we estimate a characteristic spin-up timescale around $5.5\times10^5$ yr.

\subsection{Low accretion rate: \newline transition to a ``propeller'' regime}
\label{sec:propeller}

   \begin{figure*}[!t]
   \includegraphics*[width=0.48\textwidth]{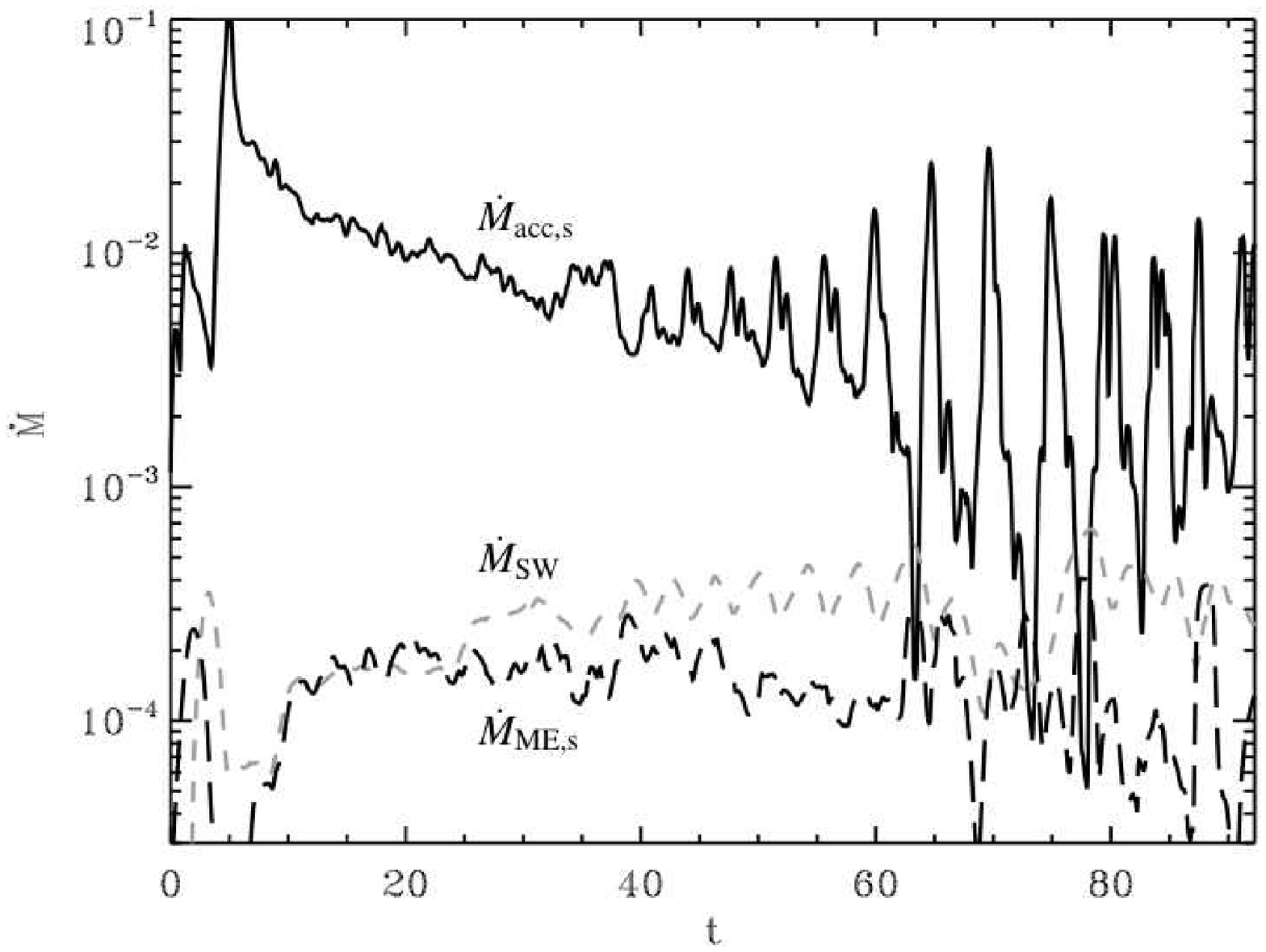}
   \hspace{0.01\textwidth}
   \includegraphics*[width=0.51\textwidth]{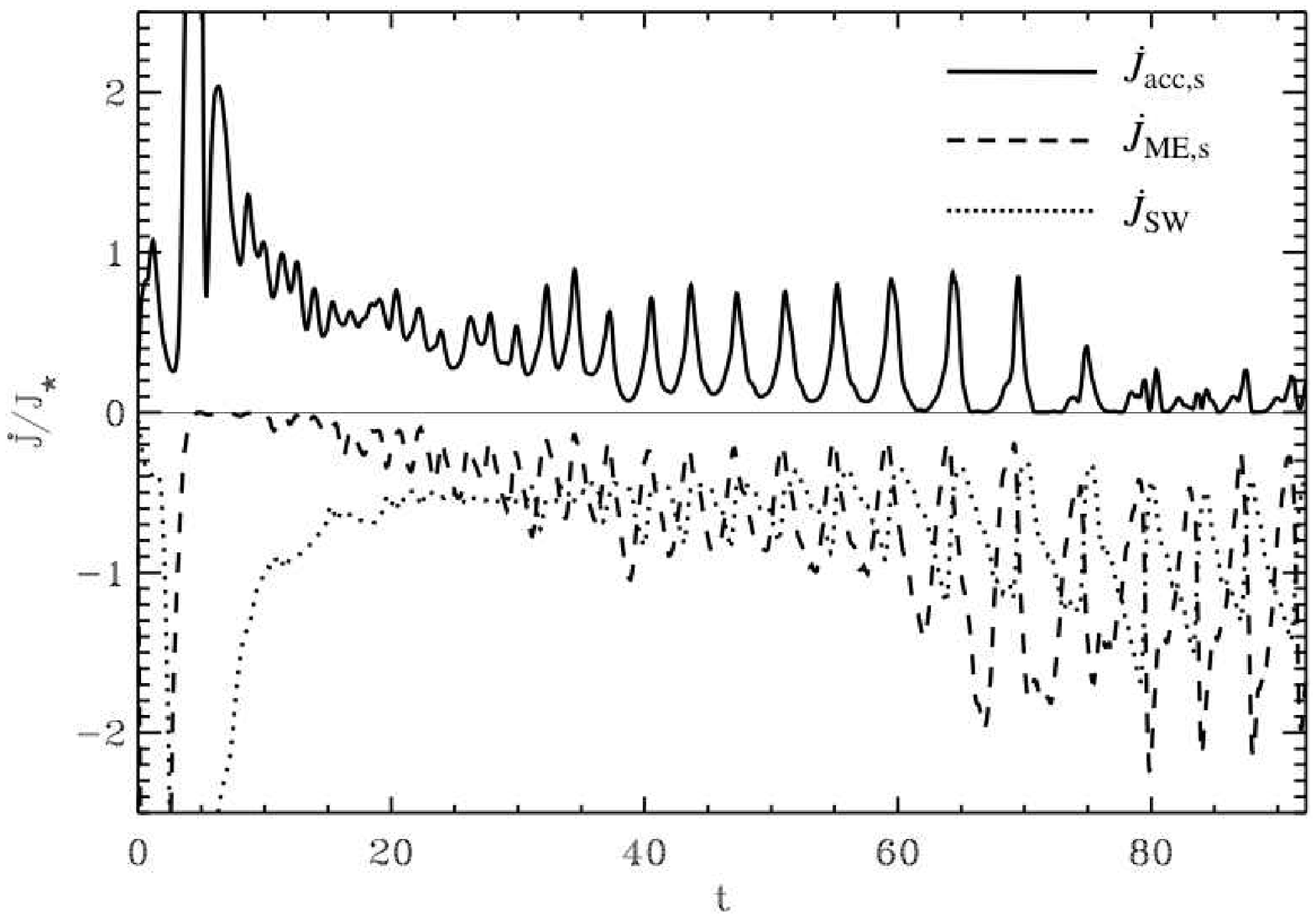}
  \caption{{\it Left panel}. Temporal evolution of different mass fluxes in case C01. Plotted are the mass accretion rate onto the stellar surface ({\it solid line}),
                    the stellar wind outflow rate ({\it dashed grey line}) and the stellar mass flux contribution to the MEs ({\it dashed line}). {\it Right panel}. Temporal evolution
                    of the torques acting onto the stellar surface, normalized to the stellar angular momentum. Plotted are the spin-up torque associated with accretion 
                    ({\it solid line}), the spin-down torque exerted along the magnetic surfaces connecting the star and the disk ({\it dashed line}) and the spin-down
                    torque due to the stellar wind ({\it dotted line}).}
  \label{fig:Alpha01rates}
  \end{figure*}  
    
Case C01 is characterized by a lower disk turbulent viscosity ($\alpha_\mathrm{v} = 0.1$) and therefore a lower mass accretion rate. Consistently
with the results of cases C03 and C1, in case C01 the viscous torque likely supports an accretion rate which is
smaller than the one determined by the inner magnetospheric torques. Therefore, the disk tends to empty, reducing its surface density and
pressure, thus decreasing the $\alpha$ torque and its accretion rate. Besides, the $\alpha$-disk model that we employed as
initial condition presents a large-scale meridional circulation pattern for low $\alpha_\mathrm{v}$ values, with the disk accreting along the surface layers 
only and excreting along the midplane. Specially for low $\alpha_\mathrm{v}$ values, this model becomes very sensitive to local changes of thermal 
pressure gradients and its accretion rate is difficult to control. 
  
\begin{figure*}[!t]
  \centering
  \includegraphics[width=17cm]{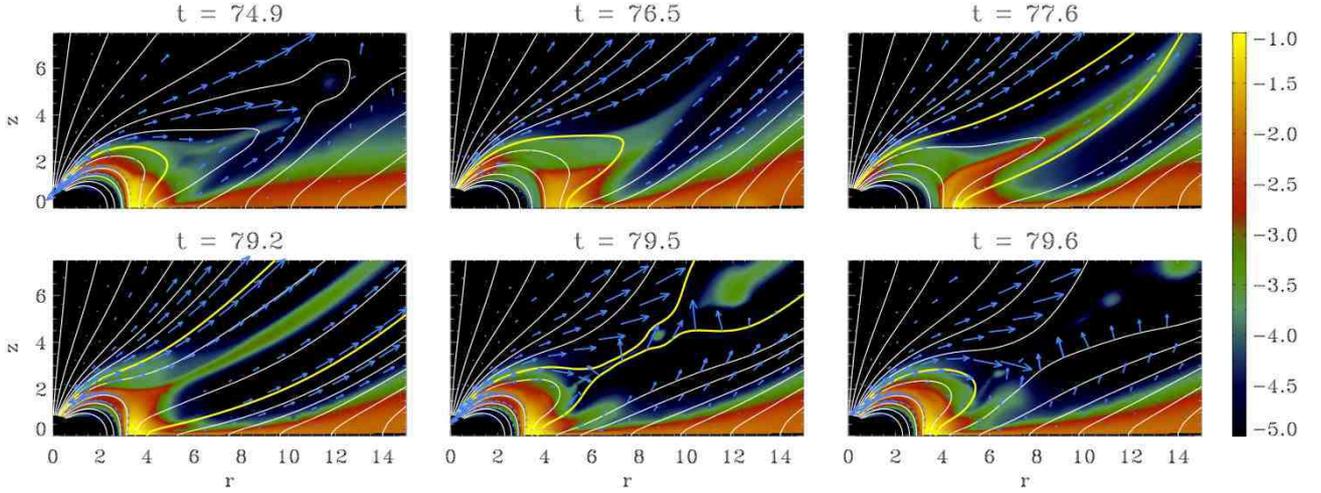}
   \caption{Time evolution of an accretion cycle during the propeller phase of case C01. We show logarithmic density maps with sample field lines (solid lines) and speed vectors (blue arrows)
                   superimposed. The yellow solid line follows the evolution of a single magnetic surface, showing clearly the periodicity of the accretion-ejection cycles. Time is given in units of the
                   stellar rotation period.}
   \label{fig:Alpha01cycle}
  \end{figure*}

Due to these effects, the disk accretion rate in this case is rather low and slowly decreases in time. This is clearly visible in the left
panel of Fig. \ref{fig:Alpha01rates}, where, after a strong initial transient peak, the accretion rate diminishes steadily, subsequently it starts to oscillate and, 
at last ($t >60$), it becomes highly unsteady and intermittent, varying periodically between relatively high and extremely low values. 
The system evolution during one of these cycles is depicted in Fig. \ref{fig:Alpha01cycle}, which clearly shows how the accretion cycles correspond
to a periodic oscillation of the truncation radius.
      
The low-accretion phases correspond to the truncation radius moving close to the Keplerian corotation radius: 
in this situation it is hard to form the accretion funnels, since the centrifugal barrier raised by the rotating magnetosphere prevents the disk material 
from falling towards the star and an extra thermal pressure gradient would be required to cross the barrier  \citep{Koldoba02}.
As extensively discussed by \citet{Bessolaz08}, the disk truncation is primarily determined by a pressure equilibrium between the disk and the magnetosphere.
Therefore in case C01, as the accretion rate decreases, the disk reduces its push against the magnetosphere and the truncation radius progressively
moves outwards closer to corotation. A consequence of the disk being truncated beyond the corotation radius is that, since in this region a Keplerian disk 
rotates slower than the star, the stellar rotation tries to increase (``propel'') the disk angular speed in the region directly connected to the star, hence
the appellative ``propeller'' regime \citep{Illarionov75, Ustyugova06}. Notice that, in our simulations, a propeller effect can be present even below the Keplerian corotation radius, 
whenever the disk rotation becomes sub-stellar, see for example Fig. \ref{fig:Comp03}.
Despite the presence of a propeller effect, the angular momentum extracted from the star is likely to be energetically insufficient to gravitationally unbind 
all accreting matter \citep{Spruit93}. Besides, as the disk moves farther from the star, the stellar magnetic field weakens and the magnetosphere gets more 
and more inflated, losing part of its connection with the disk. As the field lines open up, a massive magnetospheric outflow is launched along the opening 
magnetic surfaces. This ejections are analogous to the MEs described in the previous Sections; in a propeller phase, they
can be even stronger and can extract an important fraction of the angular momentum of  the disk. Because of the torque exerted by these strong
ejections, the disk temporarily increases its accretion rate and its drive against the stellar magnetosphere. The truncation radius can therefore move closer
to the star, the funnel flows are fueled again and the accretion rate onto the surface of the star increases. At the same time, the inflated field lines along
which the MEs have been launched tend to close again and quench the outflow (see the lower central panel in Fig. \ref{fig:Alpha01cycle}). 
Instead of losing angular momentum to the outflow, the disk can now acquire it from the star along the field lines that are connected to the star, since, 
due to the action of the MEs, the disk rotation is at least in part sub-stellar, analogously to case C03. The disk therefore decreases 
its drive against the compressed magnetosphere which pushes back the disk towards the corotation region and the cycle repeats. Notice that the disk can 
move outwards not only if it becomes super-Keplerian, but, since during the accretion phase the closed magnetosphere is strongly compressed 
(see for example Fig. 8 in \citetalias{ZanFer09}), the poloidal magnetic pressure can effectively push the disk outwards whenever the latter reduces its thrust. 

These accretion/ejection cycles have been observed in other numerical works \citep[e.g.][]{Goodson99b,Ustyugova06}: notice that the range of timescales and amplitudes of the oscillations
observed in these papers can be simply due to different parameters of the models, i.e. stellar rotation period, magnetic field intensity, disk accretion rate.
The accretion cycles of a disk truncated close to corotation have been predicted also by \citet{Spruit93} \citep[see also][]{Dangelo10}. They showed how an accretion disk truncated beyond
the Keplerian corotation radius can possibly readjust its temperature and density structure so that the viscous stresses get rid of the excess angular momentum 
coming from the star and the accretion flow can cross the corotation radius and cyclically form the accretion funnels. The characteristic period of their cycles
is obviously  associated with the viscous accretion timescale in the corotation region. In our simulation the timescale of the accretion cycles is much shorter:
since the oscillations are mainly driven by the torque exerted by the MEs, the typical period is of the order of a few Keplerian orbits around the corotation radius.
      
The torques exerted onto the star are plotted in the right panel of Fig. \ref{fig:Alpha01rates}. Clearly, during the ``propeller'' phases characterized by an
extremely low accretion rate, there is no spin-up torque associated with accretion. Besides, a strong spin-down torque is exerted along the field lines connected 
to the disk. Because of the large amount of angular momentum extracted by the magnetospheric ejections, both the accretion disk and the MEs rotate slower than
the central star, even in the sub-corotation region. Therefore the star can be efficiently spun-down along the magnetic surfaces directly connected with 
the disk and the MEs.
  
During the high accretion phases, the disk can deposit its angular momentum along the funnel flows, exerting a small but noticeable spin-up torque.
The spin-down torque exerted along the closed magnetosphere is reduced, since the MEs, which are the main responsible for this torque,
are weaker during the accretion phases: nevertheless, it is still possible to balance the torque due to accretion. The accreting phases of case C01 
resemble, at least qualitatively, the steadily accreting case C03.
  
Finally, it is important to notice that also in case C01 a strong stellar wind is present, exerting an important spin-down torque onto the star.
Table \ref{table:fluxes} shows that in this case the stellar wind can exploit a larger amount of open stellar flux that becomes even more important
during the non-accreting phases, when the stellar wind torque seems to become even stronger.
This is obviously not consistent with having a stellar wind fueled by the accretion power: in case C01 the stellar wind would require a considerable driving
power, as well as during the phases during which the disk is not accreting. 
On the other hand, in this case the role played by the stellar wind can be neglected, since the
torque exerted by the star-disk-MEs interaction is largely sufficient to brake the stellar rotation. 
Even neglecting the stellar wind torque, we can estimate that, on average, the characteristic spin-down timescale in case C01 is 
approximately equal to $8\times10^5$ yr.

\section{Discussion}
\label{sec:discussion}

In this Section we compare our findings with works that have described related scenarios and addressed similar issues. 
First of all, we point out that phenomena qualitatively similar to the MEs have been observed in a number of simulations of magnetic star-disk interaction: 
in spite of differences in the interpretation of the results, we have the feeling that these numerical experiments show essentially the same process.
For example, \citet{Hayashi96} observed a plasmoid ejection associated with the expansion of the magnetosphere and associated the reconnection
episodes to the X-ray flares observed in young stars. A series of papers \citep{Goodson97, Goodson99a, Goodson99b, Matt02} proposed 
the ``episodic magnetospheric inflation'' mechanism \citep[EMI,][]{Matt03} to explain the origin of jets from young stars: this is probably due to the fact that, 
besides the presence of uncollimated outflows commonly related to this phenomenon, these simulations show the formation of a dense axial collimated jet, 
fueled during the inflation phases by mass coming from the accretion  streams and focused around the rotation axis by the expansion of the closed magnetosphere.
While their description of this mechanism is very phenomenological, this dense collimated axial feature has not been observed in our numerical models. 
Additionally, the periodic behavior of the mass accretion rate suggests that these solutions might be in a  ``propeller'' regime. More recently, \citet{Romanova09} 
developed both axisymmetric and three dimensional non-axisymmetric simulations that describe the formation of uncollimated outflows (``conical winds'') coming 
from the  boundary of the closed magnetosphere, both for slowly and fast rotating stars.
Despite clear similarities between these results and the scenario proposed in this paper, we have performed a much more detailed analysis of the dynamics and the 
energetics of MEs, which allowed us to obtain precise results about how they can affect the angular momentum of the star-disk system. For example, we showed how
MEs can extract angular momentum both from the disk and the star so that their action can be compared to that of X-Winds on one hand and stellar winds on the other.

Similarly to X-Winds, we showed that MEs can extract a relevant fraction of the disk angular momentum in order to reduce the spin-up torque exerted
by accretion. Actually, the X-Wind model represents an extreme solution in which all the angular momentum carried by the accretion flow is
extracted from the disk, so as to cancel at least the accretion torque. For the MEs we estimated a lower efficiency, e.g. $53\%$ in the fiducial C03 case.
Due to the lower torque efficiency, the energy needed to drive MEs  is compatible with the mechanical power released by the accretion
flow, while the latter is likely to be insufficient to drive an efficient X-Wind \citep{Ferreira12}. 
In addition, the MEs can also extract angular momentum directly from the star, similarly to stellar winds. The main difference is that, while the spin-down torque
exerted by stellar winds strongly depends on the mass flux of the wind \citep{Matt08a}, the torque of the MEs is due to a differential rotation effect between
the star and the material ejected from the disk by MEs. Therefore, MEs do not seem to suffer the energetic problems associated with the mass-loading of the
stellar winds. Notice that the idea of having an outflow mass-loaded from the accretion disk capable of extracting angular momentum from the central star is 
close to the Reconnection X-Wind model \citep{Ferreira00}, even if this scenario had been envisaged for a different magnetic configuration.

An important difference between MEs and stationary winds is represented by their asymptotic behavior. 
While the large-scale acceleration and collimation properties of steady outflows,
whether coming from the star or the disk, depend on the global distribution of their poloidal magnetic flux and electric current, 
we showed that MEs rapidly disconnect from the central region of the disk-star system due to reconnection events and propagate ballistically 
afterwards, without accelerating any further.
Flowing in between a stellar wind and a disk wind (see Fig. \ref{fig:globalview}), the confinement of the MEs depends on the 
collimating/decollimating behavior of the other two outflows (see discussion in Appendix \ref{sec:current}).

A final comparison can be made between the poloidal magnetic configuration found in our simulations and the one expected from the X-Wind model, 
at least in its ``dipolar'' formulation \citep[see e.g.][]{Ostriker95}. The amount of magnetic flux contained in the magnetic cavity is rather consistent with our
results ($\Phi_\mathrm{MC}$ in Table \ref{table:fluxes}). The X-Wind model also predicts that the remaining flux \citep[see Eq. (11) in][]{Mohanty08} 
is trapped in the disk in a small region around the corotation radius and  is evenly shared among the accretion funnels ($\Phi_\mathrm{AC}$), 
the open flux ($\Phi_\mathrm{SW}$) and the remaining closed field lines which do not accrete ($\Phi_\mathrm{ME}+\Phi_\mathrm{CD}-\Phi_\mathrm{AC}$).
In our simulations the flux repartition can noticeably vary from case to case but it can be actually similar to the X-Wind distribution in cases close to the spin equilibrium 
(e.g. during the accretion phases of case C01). Three important differences can be pointed out: (1) in our simulations the magnetic flux involved in the star-disk interaction
is distributed over a sizeable region of the disk, as shown by the anchoring radii of the different magnetic surfaces given in Table \ref{table:fluxes}; 
(2) there is no magnetically ``dead zone'', namely a magnetic zone with neither mass nor angular momentum exchanges, 
as proposed in the X-wind scenario: in contrast, we obtain an extremely active, outbursting zone where MEs take place; (3) the steadily 
open magnetic surfaces threading the disk, where the X-Wind should be accelerated, play a marginal role in our solutions.

In the following subsection we are going to compare more quantitatively the efficiency of the spin-down torques found in our simulations with the outcome of other
scenarios proposed in the literature.

\subsection{Zero-torque condition}

By progressively lowering the disk accretion rate, the three simulations analyzed in the paper show a transition from a strong spin-up (case C1) to an
efficient spin-down state (case C01). 
This trend corresponds also to a different repartition of the stellar magnetic flux among the spin-down/spin-up phenomena (see Table \ref{table:fluxes}),
where the magnetic flux associated with the accretion spin-up torque ($\Phi_\mathrm{AC}$ and $\Phi_\mathrm{CD}$), 
decreases at the expenses of the spin-down mechanisms ($\Phi_\mathrm{ME}$  and $\Phi_\mathrm{SW}$).
In case C03 the torques exerted by the MEs, with a weaker but non-negligible contribution from a light stellar wind, 
are able to balance the accretion torque, so that the net total torque is approximately zero (actually slightly negative in this specific case).  Correspondingly,
the truncation radius moves from $R_\mathrm{t} \approx 0.45 R_\mathrm{co}$ (case C1) to $R_\mathrm{t} \approx 0.8 R_\mathrm{co}$ (case C01), with
the zero-torque configuration approximately located at $R_\mathrm{t} \approx 0.6 R_\mathrm{co}$. Besides, the simulations confirmed the results of
\citet{Bessolaz08} and of \citetalias{ZanFer09}, showing that the Alfv\'en radius for a spherical free-fall collapse \citep{Elsner77}, namely
\begin{equation}
R_\mathrm{A} = \left( \frac{B_\star^4 R_\star^{12}}{GM_\star \dot{M}_\mathrm{acc}^2} \right)^{1/7} \; ,
\end{equation}
is a good parametrization of the truncation radius. In our simulations we have found $R_\mathrm{t} \approx 0.4 R_\mathrm{A}$. Therefore the zero-torque state corresponds
roughly to a situation where $R_\mathrm{co} \approx 0.67 R_\mathrm{A}$. A proportionality relationship between $R_\mathrm{co}$ and $R_\mathrm{A}$ has been
used in different scenarios to estimate the stellar rotation period corresponding to a zero-torque situation:
\begin{eqnarray}
P_\mathrm{eq}  &=&  10.3\, K \, \left( \frac{B_\star}{1\, \mathrm{kG}}\right)^{6/7}  \left( \frac{R_\star}{2 \, R_\odot}\right)^{18/7}  \left( \frac{M_\star}{0.5 \, M_\odot}\right)^{-5/7} \nonumber \\
                             & & \times \left( \frac{\dot{M}_\mathrm{acc}}{10^{-8} \, M_\odot \, \mathrm{yr}^{-1}}\right)^{-3/7}  \; \mathrm{days} \; ,
\label{eq:peqdisk}                             
\end{eqnarray}
where $K = (R_\mathrm{co}/R_\mathrm{A})^{3/2}$. Obviously, the longer the equilibrium rotation period, the more efficient the spin-down mechanism.
We already showed that our simulations suggest a value $K \approx 0.54$. We can compare this result with the outcome of other popular scenarios.

Applying the classical \citeauthor{Ghosh79} scenario to the case of T Tauri stars, \citet{Konigl91} used a value $K = 0.87$. On the other hand, the corrections brought to the
\citeauthor{Ghosh79} model by \citet{Matt05a} provide an upper limit $K < 0.3$, keeping in mind that, according to \citetalias{ZanFer09}, 
the \citeauthor{Ghosh79} spin-down torque is likely to be even weaker than the \citet{Matt05a} estimate. 
In the case of the X-Wind model, \citet{Ostriker95} employed a value $K = 0.89$, even if we recall that many properties of the scenario are based on ad-hoc assumptions and
are not the result of a detailed dynamical calculation. 
The results  of the time-dependent axisymmetric simulations by \citet{Long05} have been recently re-examined, suggesting a value $K=0.52$ (M. Romanova, priv. comm.),
in good agreement with our findings.

For the accretion powered stellar wind scenario it is possible to derive an expression analogous to Eq. (\ref{eq:peqdisk}) by equating the spin-down torque of a stellar wind
$\left(\dot{J}_\mathrm{SW} = \dot{M}_\mathrm{SW} \overline{r}_\mathrm{A}^2 \Omega_\star\right)$ with the accretion torque of the disk $\left(\dot{J}_\mathrm{acc} \propto
\dot{M}_\mathrm{acc} \sqrt{GM_\star R_\mathrm{t}}\right)$. Using Eq. (12) from \citet{Matt08a} to determine the average Alfv\'en radius $\overline{r}_\mathrm{A}$ and 
assuming the accretion torque found in our case E1 since it is not affected by the presence of the MEs 
$\left(\dot{J}_\mathrm{acc}  = 1.2 \dot{M}_\mathrm{acc} \sqrt{GM_\star R_\mathrm{t}}\; \mathrm{ with } \; R_\mathrm{t} = 0.4 R_\mathrm{A}\right)$, we obtain 
\begin{eqnarray}
P_\mathrm{eq}  &\approx&  5.8\,  \left( \frac{B_\star}{1\, \mathrm{kG}}\right)^{0.61}  \left( \frac{R_\star}{2 \, R_\odot}\right)^{2.26}  \left( \frac{M_\star}{0.5 \, M_\odot}\right)^{-0.65} \nonumber \\
                             & & \times \left( \frac{\dot{M}_\mathrm{acc}}{10^{-8} \, M_\odot \, \mathrm{yr}^{-1}}\right)^{-0.3} \, \left( \frac{\dot{M}_\mathrm{SW}/\dot{M}_\mathrm{acc}}{0.1}\right)^{0.55} \; \mathrm{days} \; ,
\label{eq:peqwind}                                 
\end{eqnarray}
similarly to Eq. (17) in \citet{Matt08b}. This equation clearly shows that an ejection efficiency $\dot{M}_\mathrm{SW}/\dot{M}_\mathrm{acc} \sim 0.1$ is needed to 
give a characteristic rotation period comparable to the other scenarios, as discussed in \citet{Matt12}.

Since Eq. (\ref{eq:peqdisk}) and (\ref{eq:peqwind}) describe an equilibrium between accretion and spin-down torques, the same zero-torque condition 
(i.e. the same $P_\mathrm{eq}$) can be obtained by varying accordingly the mass accretion rate, proportional to the accretion torque, and the stellar magnetic 
field intensity, mainly related to the spin-down torque. This implies that stars with a comparatively lower accretion rate need a weaker dipolar field to balance
the accretion torque.

\subsection{Stellar contraction and spin-down}

The zero-torque condition discussed in the previous Section is rather challenging to obtain for all the discussed models, but it is obviously not sufficient to keep constant the rotation
period of a contracting star. The time derivative of the stellar angular velocity
\begin{equation}
I_{\star}\frac{\mathrm{d} \Omega_{\star}}{\mathrm{d} t} = \frac{\mathrm{d} J_\star}{\mathrm{d} t} - J_{\star}\left(\frac{\dot{M}_{\star}}{M_{\star}} +  
\frac{2\dot{R}_\star}{R_{\star}}\right) \; ,
\label{eq:omegadot}
\end{equation}
depends also on the temporal evolution of the stellar mass and radius. The right hand side of Eq. (\ref{eq:omegadot}) defines three timescales: the term
$\dot{M}_\star/M_\star \sim 1/t_\mathrm{acc}$ is associated with the accretion timescale, typically $t_\mathrm{acc} \sim 10^7 - 10^{10}$ yr for a CTTS; the term
$\dot{R}_\star/R_\star \sim -1/t_\mathrm{KH}$ is associated with the the Kelvin-Helmholtz contraction timescale:
\begin{equation}
t_\mathrm{KH} = 1.8 \times 10^6 \; \left( \frac{R_\star}{2 \, R_\odot}\right)^{-3}  \left( \frac{M_\star}{0.5 \, M_\odot}\right)^{2}  \left( \frac{T_\mathrm{eff}}{4000 \, \mathrm{K}}\right)^{-4} \; \mathrm{yr} \; .
\end{equation}
Clearly, it is necessary to exert a net negative torque onto the star with a characteristic spin-down timescale $J_\star/\dot{J}_\star < 0$ at least comparable to the contraction timescale to
maintain a steady stellar rotation period. In our simulations this condition is clearly satisfied during the propeller phases of case C01. This regime occurs whenever the disk is truncated
close enough to the corotation radius ($R_\mathrm{t} \gtrsim 0.8 R_\mathrm{co}$ in our simulations). Notice however that this condition is necessary but not sufficient to have a spin-down
torque strong enough to balance the stellar contraction. While different $B_\star - \dot{M}_\mathrm{acc}$ combinations can satisfy the condition $R_\mathrm{t} \approx 0.8 
R_\mathrm{co}$, as in the case of a null torque configuration, a strong magnetic field is necessary to provide a spin-down torque able to balance the stellar contraction. This is clearly shown
by Eq. (\ref{eq:tnorm}), used to normalize the torques plotted in Fig. \ref{fig:Alpha03torques}, \ref{fig:Alpha1torques} and \ref{fig:Alpha01rates}. In case C01 we must assume
a strong dipolar component, typically of the order of the kG, to obtain a short enough spin-down timescale. This result shows that, even if stars characterized by a low accretion rate
(e.g. $< 10^{-9} M_\odot \; \mathrm{\mbox{yr}}^{-1}$) need a dipolar field weaker than a kG to cancel the accretion torque, they still need a kG dipole to balance their contraction.

\section{Summary and conclusions}
\label{sec:summary}

In this paper we presented axisymmetric MHD time-dependent simulations of the interaction of a dipolar stellar magnetosphere with a surrounding viscous 
and resistive accretion disk (using $\alpha$ prescriptions). 
We assumed a magnetic coupling strong enough (i.e. a resistivity sufficiently low) so that the buildup of the toroidal field magnetic pressure due to the star-disk differential rotation
inflates and opens up the dipolar structure close to the central star and the truncation region. In particular, the strong coupling prevents the closed magnetosphere to extend beyond 
the Keplerian corotation radius, so that the \citeauthor{Ghosh79} spin-down model, studied in \citetalias{ZanFer09}, cannot be applied anymore.
On the other hand, the simulations showed that magnetospheric ejections naturally arise because of the process of expansion and reconnection of the magnetospheric field lines 
connected to the disk. We extensively studied the dynamical properties of these ejections with a special focus on their impact on the angular momentum of the star-disk system.
At the same time we have included in our models the spin-down torque exerted by a stellar wind. 
We here summarize the main results of our numerical experiments:

\begin{enumerate}

\item MEs can exchange mass, energy and angular momentum both with the star and the disk. Their inertia and mass content is largely dominated by the material 
          loaded from the accretion disk. If the mass load rotates slower than the star the ejections can be powered by both the stellar and disk rotation, as in a huge
          magnetic slingshot. 

\item MEs cannot explain the jet phenomenon in T Tauri stars: (1) their terminal speed is unlikely to be higher than the gravitational escape speed; (2) after they 
          disconnect from the star-disk system they propagate ballistically as uncollimated plasmoids. Their confinement depends on the collimation properties of the 
          outflows between which they propagate, stellar and disk winds.

\item MEs crucially contribute to control the stellar rotation period. On one hand, they efficiently extract angular momentum from the disk close to the truncation
          region so that the spin-up accretion torque is sensibly reduced. On the other hand, if the torque exerted onto the disk is strong enough so that the ejected
          plasma rotates slower than the star, MEs can extract angular momentum directly from the star thanks to a differential rotation effect (slingshot effect). 
          We found a balance between the accretion torque and the spin-down torque for $R_\mathrm{t} \approx 0.6 R_\mathrm{co}$.
          
\item The efficiency of the spin-down torque of MEs is comparable to other scenarios proposed in the literature, keeping in mind that: the \citet{Ghosh79}
          spin-down torque is highly overestimated \citep[see][]{Matt05a,ZanFer09}; a fully self-consistent dynamical model of the X-Wind scenario is currently
          not available; the mass ejection efficiency of a stellar wind capable to balance at least the accretion torque is energetically very demanding.
          While the torques exerted by MEs onto the disk and the star share some similarities with the X-Wind and stellar wind models respectively, 
          MEs do not seem to be affected by the energetic limitations that concern the other two scenarios.

\item We limited the mass ejection efficiency of the stellar winds to a few percent. Consistently with the \citet{Matt08b} results, these winds are not sufficient
          to balance the spin-up due to accretion and contraction. Nevertheless, we found that for a wind mass flux around $1-2\%$ of the accretion rate, 
          the spin-down torque corresponds to $20 - 30\%$ of the accretion torque. We found that a light disk wind is launched along the open magnetic surfaces
          threading the disk but, due to the weakness of the field, the impact of this outflow on the angular momentum structure of the Keplerian disk is negligible.
          
\item In a propeller phase, when the truncation radius gets sufficiently close to corotation ($R_\mathrm{t} \gtrsim 0.8 R_\mathrm{co}$), the spin-down torque
          exerted by the disk and the MEs can even balance the spin-up due to contraction. On the other hand, during the propeller phases the accretion becomes
          intermittent on a dynamical timescale, corresponding to a few rotation periods of the star. Even if this effect is enhanced by the axial symmetry of our
          models, there is no observational evidence of such a behavior.

\end{enumerate}

We want to conclude by pointing out a limit that affects all the scenarios discussed in this work, including the model presented in this paper. All the scenarios
are based on axisymmetric models of purely dipolar stellar magnetospheres and come to the conclusion that, for typical slowly rotating CTTS, a dipolar 
component around $\sim 1$ kG is needed to balance, at least, the spin-up torque due to accretion. While this has been observed, for example, in the case 
of AA Tau \citep{Donati10b}, such a strong dipolar component does not seem to be a common feature among T Tauri stars \citep{Donati07, Donati10a, 
Donati11a, Donati11b, Donati11c, Hussain09}. While a larger sample of magnetic field measurements is clearly needed, stellar torque models must start 
to consider more realistic magnetic field configurations, including non axisymmetric and multipolar components \citep[see e.g.][]{Long11, Romanova11, Vidotto11}.

\begin{acknowledgements}
The authors thank the referee for his  insightful comments, J\'er\^ome Bouvier, Sean Matt, Marina Romanova for helpful discussions and Sylvie Cabrit for
suggesting the slingshot analogy. C.Z. acknowledges support from the Marie Curie Action ``European Reintegration Grants'' under contract PERG05-GA-2009-247415.
\end{acknowledgements}

  \begin{figure*}[!t]
  \centering
  \includegraphics[width=17cm]{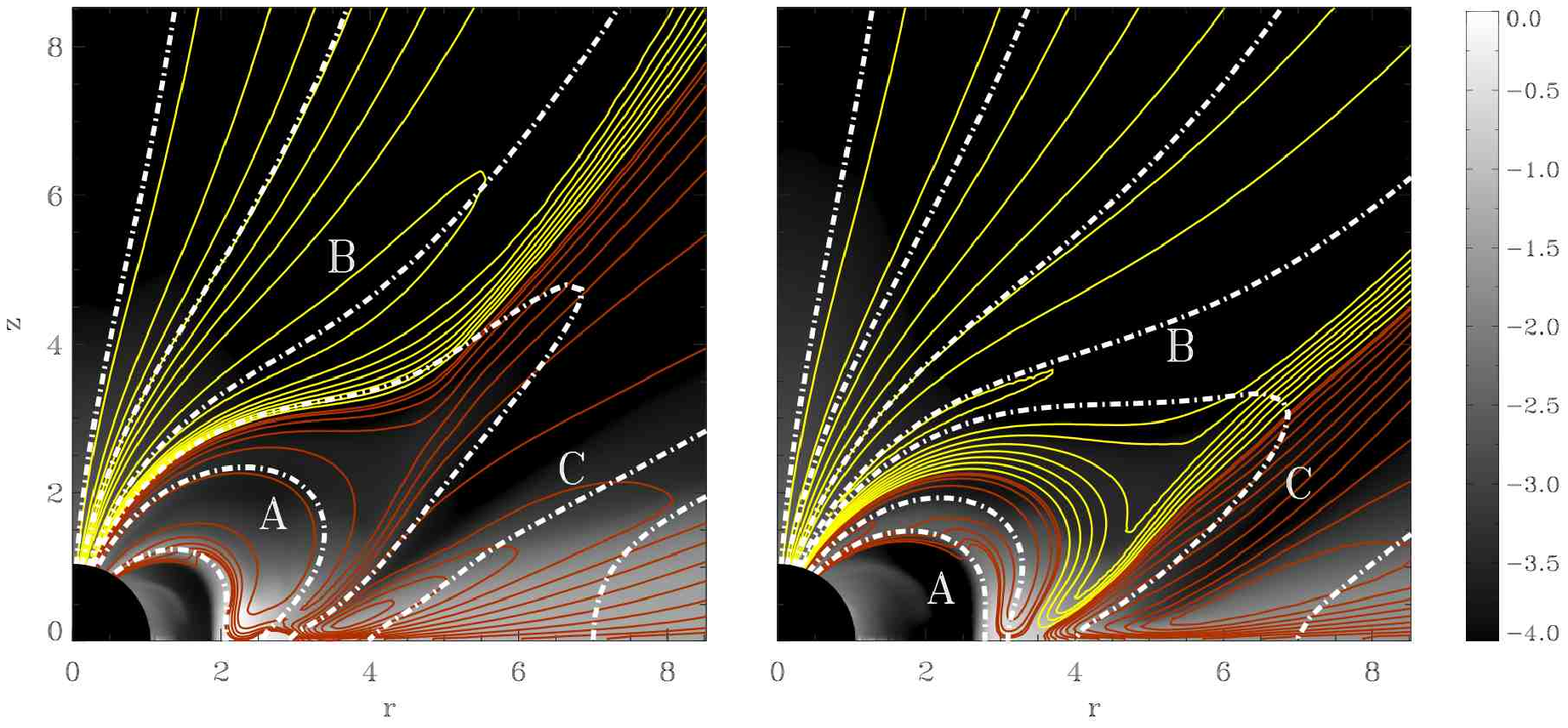}
   \caption{Poloidal current circuits flowing in the star-disk system in cases C1 (left panel) and C03 (right panel). 
                   Dark (red) circuits circulate clockwise along isosurfaces of $rB_\phi > 0$. 
                   Yellow (light) circuits circulate counterclockwise along isosurfaces of $rB_\phi <  0$. Current circuits are superimposed to a logarithmic
                   density map.  The figures have been obtained by time averaging over 54 stellar rotation periods.}
   \label{fig:currents}
  \end{figure*} 
  
   \begin{figure*}[!t]
  \centering
  \includegraphics[width=17cm]{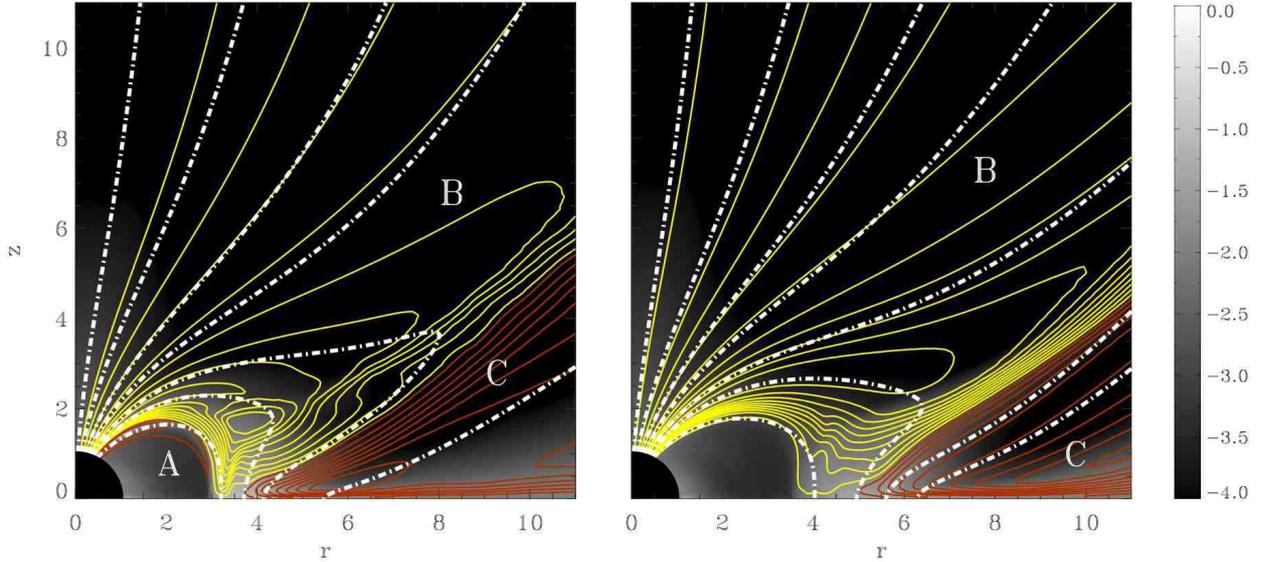}
   \caption{Poloidal current circuits flowing in the star-disk system in case C01. Color codes are the same as in Fig. \ref{fig:currents}. The left panel
                   refers to the accretion phases, while the right panel represents the strictly propeller phases. The left (right) panel has been obtained by
                   averaging in time current and density during the maxima (minima) of the accretion rate for $t > 60$, see Fig. \ref{fig:Alpha01rates}. }
                   
   \label{fig:currentspropel}
  \end{figure*} 
  
\appendix 
\section{Global magnetic coupling}
\label{sec:current}
  
Since the star-disk interaction is mainly controlled by magnetic stresses, the best way to have a general view of the magnetic coupling between the different
parts of the system, the accretion disk, the star and the outflows, is to look at the poloidal electric currents $\vec{J}_\mathrm{p}$, circulating along the 
$r B_\phi =$ const. isosurfaces. The poloidal component of the Lorentz force $\vec{J}_\mathrm{p} \times \vec{B}_\phi$ is perpendicular to these surfaces, 
while the toroidal force $\vec{J}_\mathrm{p} \times \vec{B}_\mathrm{p}$ is obviously proportional to the magnetic torque.  Besides, in the region close to the 
star, the dominant components of the angular momentum and energy poloidal fluxes are directed along the poloidal magnetic field and are proportional to
$\propto - B_\phi \vec{B}_\mathrm{p}$: the sign of $B_\phi$ gives therefore an information about the direction of the Poynting flux.
In Fig. \ref{fig:currents} we plot sample $r B_\phi =$ const. isosurfaces to outline the poloidal current circuits
characterizing cases C1 and C03. The two panels of Fig. \ref{fig:currentspropel} refer to the current configuration during the high accretion (left panel)
and low accretion phases (right panel) of the propeller stage ($t> 60$) of case C01.
We can distinguish three main current circuits labeled as A, B and C in the four panels. In circuits A and C the currents circulates clockwise, 
while counterclockwise in circuit B.  Different colors correspond to negative or positive values  of $B_\phi$: dark-red circuits ($B_\phi >  0$) are 
responsible for extracting angular momentum and energy from the disk, while the light-yellow circuits ($B_\phi < 0$) extract angular momentum and 
energy from the star. 

The electromotive force of A-labeled circuits is due to the star-disk differential rotation: the current flows out from the stellar surface towards the disk,
flows back along the accretion funnels and closes inside the star. This current circuit exerts a toroidal braking force inside the disk and the accretion columns 
and a spinning-up force inside the star, thus transferring angular momentum from the disk and the funnels to the star. It is therefore responsible 
for the accretion spin-up torque plotted in Fig. \ref{fig:Alpha03rates} and \ref{fig:Alpha1rates}. Notice how this circuit becomes smaller and smaller as the
accretion torque decreases going from case C1 to the high accretion phases of case C01 and completely disappears in the low accretion stages of case C01,
during which the accretion torque is completely negligible.

Circuit C brakes down the disk rotation and is responsible for the magnetic driving of outflows launched from the disk, in particular for the magnetic acceleration
of the part of the MEs mass-loaded from the disk. In the launching region of the MEs, it provides a strong vertical force which uplifts matter at the disk surface 
(see Fig. \ref{fig:forcesmej}), thus contributing to the high mass ejection efficiency of these outflows. This circuit corresponds to the innermost part of 
the butterfly-shaped current circuits characteristic of extended disk winds, see \citet{Ferreira97}.
  
Current circuit B  is associated with the energy and angular momentum extraction from the star. The current flows out from the stellar surface at mid latitudes 
and flows back to the star at higher latitudes. In Case C1 the current flows mostly along the open field lines anchored onto the stellar surface, 
therefore fueling the stellar wind. There is no spin-down circuit associated with the MEs or the disk inside the closed magnetosphere. 
On the other hand, the other cases show clearly that the spin-down circuit B couples the star with the disk, the MEs and the stellar wind. These three 
dynamical constituents can in fact extract a fraction of the  stellar angular momentum. The current flowing inside the disk, the MEs and the stellar wind 
provides in fact a $\vec{J} \times \vec{B}$ force which tries to spin-up the material rotating at sub-stellar rates, so that the star loses its angular 
momentum. The circuit B clearly identifies the parts of the disk still magnetically connected to the star which rotate slower than the latter. 

Altogether, the four panels show how the spin-up effects weaken and the spin-down action strengthens going from case C1 to case C01. 
We can actually see that in case C1 the MEs, which are efficiently extracting angular momentum from the disk, are transferring energy and angular momentum to
the star, thus providing a spin-up torque, as we saw in Section \ref{sec:highrate}. In case C03 MEs extract energy and angular momentum 
both from the star and the disk and the two fluxes converge at the cusp of the field line. In the propeller phases of case C01, MEs seem to be powered
almost exclusively by the stellar rotation. 

Finally, these figures clearly show that on a large scale MEs tend to propagate along the current sheet 
at the interface between circuits B and C. While circuit B provides a decollimating  Lorentz force, circuit C tries to collimate towards the axis. 
Therefore, since MEs propagate ballistically after they have disconnected, their collimation properties depend on the equilibrium between the 
decollimating inner field lines and the collimating outer ones, i.e. on the equilibrium between the inner stellar wind and an outer disk wind. 
A proper treatment of this issue therefore requires simulations covering larger spatial scales, properly treating all the outflows present in the system. 
For example, the large-scale simulations presented by \citet{Goodson99a} suggest that an inertial confinement due to a very thick disk and a rather dense corona 
can help focusing the closed magnetosphere expansion and the propagation of the magnetospheric outflows towards the axis \citep[see also][]{Li01}. 
\citet{Matt03} showed how a strong enough outer poloidal magnetic field anchored in the disk can confine the EMIs. More recently, \citet{Lii12} showed that, at least for
high mass accretion rates, more typical of  EXORs and FUORs, the innermost open disk field lines can be sufficiently mass loaded so that the corresponding
disk wind can collimate the inner outflows.

\begin{thebibliography}{}

\bibitem[Agapitou \& Papaloizou(2000)]{AgaPap00} Agapitou, V. \& Papaloizou, J. 2000, \mnras, 317, 273
\bibitem[Aly \& Kuijpers(1990)]{Aly90} Aly, J. J. \& Kuijpers, J. 1990, \aap, 227, 473
\bibitem[Anderson et al.(2005)]{Anderson05} Anderson, J. M., Li, Z.-Y., Krasnopolsky, R. \& Blandford, R. D. 2005, \apj, 630, 945
\bibitem[Bessolaz et al.(2008)]{Bessolaz08} Bessolaz, N., Zanni, C., Ferreira, J., Keppens, R. \& Bouvier, J. 2008, \aap, 478, 155
\bibitem[Bouvier et al.(1993)]{Bouvier93} Bouvier, J., Cabrit, S., Fern\'andez, M., Mart\'in, E. L. \& Matthews, J. M. 1993, \aap, 272, 176
\bibitem[Burrows et al.(1996)]{Burrows96} Burrows, C. J.,  Stapelfeldt, K. R.,  Watson, A. M. et al. 1996, \apj, 473, 437
\bibitem[Cabrit et al.(1990)]{Cabrit90} Cabrit, S., Edwards, S., Strom, S. E., \& Strom, K. M. 1990, \apj, 354, 687
\bibitem[Cabrit(2009)]{Cabrit09} Cabrit, S. 2009, in Protostellar Jets in Context, ed. K. Tsinganos, T. Ray \& M. Stute (Berlin: Springer), 247
\bibitem[Cai et al.(2008)]{Cai08} Cai, M. J., Shang, H., Lin, H.-H. \& Shu, F. H. 2008, \apj, 672, 489
\bibitem[Cranmer(2008)]{Cranmer08} Cranmer, S. R. 2008, \apj, 689, 316
\bibitem[Cranmer(2009)]{Cranmer09} Cranmer, S. R. 2009, \apj, 706, 824
\bibitem[D'Angelo \& Spruit(2010)]{Dangelo10} D'Angelo, C. \& Spruit, H. C. 2010, \mnras, 406, 1208
\bibitem[Donati et al.(2007)]{Donati07} Donati., J.-F., Jardine, M. M., Gregory, S. G. et al. 2007, \mnras, 380, 1297
\bibitem[Donati et al.(2010a)]{Donati10a} Donati., J.-F., Skelly, M. B., Bouvier, J. et al. 2010a, \mnras, 402, 1426 
\bibitem[Donati et al.(2010b)]{Donati10b} Donati., J.-F., Skelly, M. B., Bouvier, J. et al. 2010b, \mnras, 409, 1347
\bibitem[Donati et al.(2011a)]{Donati11a} Donati., J.-F., Bouvier, J. Walter, F. M. et al. 2011a, \mnras, 412, 2454
\bibitem[Donati et al.(2011b)]{Donati11b} Donati., J.-F., Gregory, S. G., Alencar, S. et al. 2011b, \mnras, 417, 472
\bibitem[Donati et al.(2011c)]{Donati11c} Donati., J.-F., Gregory, S. G., Montmerle, T. et al. 2011c, \mnras, 417, 1747
\bibitem[Edwards et al.(1994)]{Edwards94} Edwards, S., Hartigan, P., Ghandour, L., \& Andrulis, C. 1994, \aj, 108, 1056
\bibitem[Elsner \& Lamb(1977)]{Elsner77} Elsner, R. F., \& Lamb, F. K. 1977, \apj, 215, 897
\bibitem[Ferreira(1997)]{Ferreira97} Ferreira, J. 1997, \aap, 319, 340
\bibitem[Ferreira et al.(2000)]{Ferreira00} Ferreira, J., Pelletier, G. \& Appl, S. 2000, \mnras, 312, 387
\bibitem[Ferreira et al.(2006)]{Ferreira06} Ferreira, J., Dougados, C. \& Cabrit, S. 2006, \aap, 453, 785
\bibitem[Ferreira \& Casse(2012)]{Ferreira12} Ferreira, J. \& Casse, F. 2012, \mnras, in press, \texttt{arXiv:1209.3871 [astro-ph.SR]}
\bibitem[Ghosh \& Lamb(1979)]{Ghosh79} Ghosh, P. \& Lamb, F. K. 1979, \apj, 234, 296
\bibitem[G\'omez de Castro \& von Rekowski(2011)]{Gomez11} G\'omez de Castro, A. I. \& von Rekowski, B. 2011, \mnras, 411, 849
\bibitem[Goodson et al.(1997)]{Goodson97} Goodson, A. P., Winglee, R. M. \& B\"ohm, K.-H. 1997, \apj, 489, 199 
\bibitem[Goodson et al.(1999a)]{Goodson99a} Goodson, A. P., B\"ohm, K.-H. \& Winglee, R. M. 1999a, \apj, 524, 142
\bibitem[Goodson et al.(1999b)]{Goodson99b} Goodson, A. P. \& Winglee, R. M. 1999b, \apj, 524, 159
\bibitem[Hartmann et al.(1998)]{Hartmann98} Hartmann, L., Calvet, N., Gullbring, E. \& D'Alessio, P. 1998, \apj, 495, 385
\bibitem[Hartmann(2002)]{Hartmann02} Hartmann, L. 2002, \apj, 566, 29
\bibitem[Hartmann(2009)]{Hartmann09} Hartmann, L. 2009, in Protostellar Jets in Context, ed. K. Tsinganos, T. Ray \& M. Stute (Berlin: Springer), 23
\bibitem[Hayashi et al.(1996)]{Hayashi96} Hayashi, M. R., Shibata, K. \& Matsumoto, R. 1996, \apj 468, 37
\bibitem[Hussain et al.(2009)]{Hussain09} Hussain, G. A. J., Collier Cameron, A., Jardine, M. M. et al. 2009, \mnras, 398, 189
\bibitem[Illarionov \& Sunyaev(1975)]{Illarionov75}  Illarionov, A. F. \& Sunyaev, R. A. 1975, \aap, 39, 185
\bibitem[Irwin \& Bouvier(2009)]{Irwin09} Irwin, J. \& Bouvier, J. 2009, in IAU Symp. 258, The Ages of Stars, ed. E. E.
                                                  Mamajek, D. R. Soderblom, \& R. F. G.Wyse (Cambridge: Cambridge Univ. Press), 363
\bibitem[Johns-Krull(2007)]{JK07} Johns-Krull, C. M. 2007, \apj, 664, 975
\bibitem[Klu\'zniak \& Kita(2000)]{KluzniakKita00} Klu\'zniak, W., \& Kita, D. 2000, ArXiv e-prints, \texttt{arXiv:astro-ph/0006266}
\bibitem[Koldoba et al.(2002)]{Koldoba02} Koldoba, A. V., Lovelace, R. V. E., Ustyugova, G. V. \& Romanova, M. M. 2002, \apj, 123, 2019
\bibitem[K\"onigl(1991)]{Konigl91} K\"onigl. A.  1991, \apj, 370, 39
\bibitem[Li et al.(2001)]{Li01} Li, H., Lovelace, R. V. E., Finn, J. M. \& Colgate, S. A. 2001, \apj, 561, 915
\bibitem[Lii et al.(2012)]{Lii12} Lii, P. ,Romanova, M. \& Lovelace, R. V. E. 2012, \mnras, 420, 2020
\bibitem[Long et al.(2005)]{Long05} Long, M., Romanova, M. M., \& Lovelace, R. V. E.  2005, \apj, 634, 1214
\bibitem[Long et al.(2011)]{Long11} 	Long, M., Romanova, M. M., Kulkarni, A. K. \& Donati, J.-F. 2011, \mnras, 413, 1061
\bibitem[Matt et al.(2002)]{Matt02} Matt, S., Goodson, A. P., Winglee, R. M. \& B\"ohm, K.-H. 2002, \apj, 574, 232
\bibitem[Matt et al.(2003)]{Matt03} Matt, S., Winglee, R. M. \& B\"ohm, K.-H. 2003, \apj, 345, 660
\bibitem[Matt \& Pudritz(2005a)]{Matt05a} Matt, S. \& Pudritz, R. E. 2005a, \mnras, 356, 167
\bibitem[Matt \& Pudritz(2005b)]{Matt05b} Matt, S. \& Pudritz, R. E. 2005b, \apj, 632, 135
\bibitem[Matt \& Pudritz(2007)]{Matt07} Matt, S. \& Pudritz, R. E. 2007, in IAU Symposium, Vol. 243, Star-Disk Interaction in Young Stars, 
            ed. J. Bouvier \& I. Appenzeller, 299
\bibitem[Matt \& Pudritz(2008a)]{Matt08a} Matt, S. \& Pudritz, R. E. 2008a, \apj, 678, 1109
\bibitem[Matt \& Pudritz(2008b)]{Matt08b} Matt, S. \& Pudritz, R. E. 2008b, \apj, 681, 391
\bibitem[Matt et al.(2012)]{Matt12} Matt, S. P., Pinz\'on, G., Greene, T. P. \& Pudritz, R. E. 2012, \apj, 745, 101
\bibitem[Mignone et al.(2007)]{Mignone07} Mignone, A., Bodo, G., Massaglia, S., et al.  2007, \apjs, 170, 228
\bibitem[Miller \& Stone(1997)]{Miller97} Miller, K. A. \& Stone, J. M. 1997, \apj, 489, 890
\bibitem[Mohanty \& Shu(2008)]{Mohanty08} Mohanty, S. \& Shu, F. H. 2008, \apj, 687, 1323
\bibitem[Najita \& Shu(1994)]{Najita94} Najita, J. R. \& Shu, F. H. 1994, \apj, 429, 808
\bibitem[Ostriker \& Shu(1995)]{Ostriker95} Ostriker, E. C., \& Shu, F. H. 1995, \apj,  447, 813
\bibitem[Regev \& Gitelman(2002)]{RegevGit02} Regev, O. \& Gitelman, L. 2002, \aap, 396, 623
\bibitem[Romanova et al.(2009)]{Romanova09} Romanova, M. M., Ustyugova, G. V., Koldoba, A. V. \& Lovelace, R. V. E. 2009, \mnras, 399, 1802
\bibitem[Romanova et al.(2011)]{Romanova11} Romanova, M. M., Long, M., Lamb, F. K., Kulkarni, A. K. \& Donati, J.-F. 2011, \mnras, 411, 915
\bibitem[Sauty et al.(2011)]{Sauty11} Sauty, C., Meliani, Z., Lima, J. J. G., Tsinganos, K., Cayatte, V. \& Globus, N. 2011, \aap, 533, 46
\bibitem[Shakura \& Sunyaev(1973)]{Shakura73} Shakura, N. I., \& Sunyaev, R. A. 1973, \aap, 24, 337
\bibitem[Shu et al.(1994)]{Shu94} Shu, F. H., Najita, J. R., Ostriker, et al. 1994, \apj, 429, 781
\bibitem[Spruit \& Taam(1993)]{Spruit93} Spruit, H. C. \& Taam, R. E. 1993, \apj, 402, 593
\bibitem[Umurhan et al.(2006)]{Umurhan06} Umurhan, O. M., Nemirovsky, A., Regev, O. \& Shaviv, G. 2006, \aap, 446, 1
\bibitem[Ustyugova et al.(2006)]{Ustyugova06} Ustyugova, G. V., Koldoba, A. V., Romanova, M. M. \& Lovelace, R. V. E. 2006, \apj, 646, 304
\bibitem[Uzdensky et al.(2002)]{UKL02} Uzdensky, D. A., K\"onigl, A. \& Litwin, C. 2002, \apj, 565, 1191
\bibitem[Vidotto et al.(2011)]{Vidotto11} Vidotto, A. A., Jardine, M., Opher, M., Donati, J. F. \& Gombosi, T. I. 2011, \mnras, 412, 351
\bibitem[Yang \& Johns-Krull(2011)]{Yang11} Yang, H. \& Johns-Krull, C. M. 2011, \apj, 729, 83
\bibitem[Zanni \& Ferreira(2009)]{ZanFer09} Zanni, C. \& Ferreira, J. 2009, \aap,  512, 1117, (Paper I)
\bibitem[Zanni \& Ferreira(2011)]{ZanFer11} Zanni, C. \& Ferreira, J. 2011, \apjl,  727L, 22

\end{thebibliography}
\end{document}